\documentclass[12pt]{article}
\usepackage{amssymb,graphicx}
\usepackage{epstopdf}
\usepackage{amsmath,amsfonts}
\usepackage{epsfig}

\def\e{{\rm e}}

\def\d{\partial}
\def\l{\left(}
\def\r{\right)}

\newcommand{\be}{\begin{equation}}
\newcommand{\ee}{\end{equation}}
\newcommand{\bea}{\begin{eqnarray}}
\newcommand{\eea}{\end{eqnarray}}
\newcommand{\bg}{\begin{gather}}
\newcommand{\eg}{\end{gather}}
\newcommand{\bseq}{\begin{subequations}}
\newcommand{\eseq}{\end{subequations}}

\renewcommand{\ln}{\mathop{\rm ln}\nolimits}

\def\half{\frac{1}{2}}

\begin{document}
\begin{flushright}
\end{flushright}
\vspace{10pt}
\begin{center}
  {\LARGE \bf The Null Energy Condition \\[0.3cm] and its violation } \\
\vspace{20pt}
V. A. Rubakov\\
\vspace{15pt}
\textit{
Institute for Nuclear Research of
         the Russian Academy of Sciences,\\  60th October Anniversary
  Prospect, 7a, 117312 Moscow, Russia;}\\
\vspace{7pt}
\textit{
Department of Particle Physics and Cosmology,\\
Physics Faculty, M. V. Lomonosov Moscow State University\\ Vorobjevy Gory,
119991, Moscow, Russia
}

    \end{center}
    \vspace{5pt}

\begin{abstract}
We give a mini-review of scalar field theories with second-derivative
Lagrangians, whose field equations are second order.
Some of these theories admit solutions violating the
Null Energy Condition and having no obvious pathologies. 
We give a few examples of using these 
theories in cosmological setting and also in the context of the
creation of a universe in the laboratory. 
\end{abstract}

\tableofcontents

\vspace{1cm}

\section{Introduction}

Among various energy conditions discussed in the context of 
General Relativity, the Null Energy Condition (NEC) plays a special role.
This condition states that the matter energy-momentum tensor 
$T_{\mu \nu}$ obeys
\be
    T_{\mu \nu} n^\mu n^\nu > 0 \; ,
\label{jan2-14-9}
\ee
for any null (light-like) vector $n^\mu$, i.e., for any vector satisfying 
$g_{\mu \nu} n^\mu n^\nu = 0$.
The reason for the NEC being particularly interesting, is twofold. 
First, the NEC is quite robust; we illustrate this point in 
Section~\ref{robust}. In fact, until rather recently the
common lore was that the NEC could not be violated in a healthy
theory, with possible exception of scalar field non-minimally
coupled to gravity~\cite{Flanagan:1996gw}. 
The developments that refuted this viewpoint are the main
emphasis of this mini-review.  

Second, the NEC is a crucial assumption of
the Penrose singularity theorem~\cite{Penrose}, valid in
General Relativity.  
The theorem assumes that (i) the NEC holds; (ii) Cauchi 
hypersurface is non-compact.
The theorem states that once there is a trapped surface in space,
then there will be a singularity in future. A trapped surface
is a closed surface on which outward-pointing light rays 
are actually converging (moving inwards). In a spherically symmetric 
situation, this means the following. Let $R$ be a coordinate that
measures the area of a sphere, $S(R) = 4\pi R^2$. Then a sphere is a trapped
surface if $R$ decreases along any future null direction; all light rays
emanating from this sphere in this sense
move towards its center. 
See Apendix A for details.
An example is a sphere inside the horizon of the Schwarzschild black
hole, or in the case of contracting, spatially flat 
homogeneous isotropic Universe, a sphere
of size greater than $|H|^{-1}$, where $H$ is the Hubble parameter.
Thus, for matter obeying the NEC, there is always a singularity
that gets formed inside a black hole horizon, and any contracting
Universe ends up in a singularity, provided its spatial curvature
is dynamically negligible (which is often the case). By time reversal,
an expanding Universe has a singularity in the past.
All this is true
in classical General Relativity; things are different in 
other classical theories of gravity, and probably very different 
in quantum gravity. 

In particular, the Penrose theorem almost forbids, within classical
General Relativity,
a bouncing Universe scenario, in which the Universe contracts at early
times, contraction terminates at some moment of time 
and the Universe
enters the expansion epoch which contitues until today.
Let us see explicitly that
 the NEC is crucial for that ban. Consider a homogeneous 
isotropic Universe with Friedmann--Lema\^itre--Robertson--Walker
 metric
\be
ds^2 = dt^2 - a^2(t) \gamma_{ij} dx^i dx^j \; ,
\label{jan2-14-1}
\ee
where $\gamma_{ij}$ is time-independent metric of unit 3-sphere
(one assigns a parameter $\kappa = +1$ to this case) or unit
3-hyperboloid ($\kappa= -1$) or Euclidean 3-dimensional space
($\kappa = 0$). Matter governing the evolution of this Universe
must  also be homogeneous and isotropic, meaning that the only
non-vanishing components of energy-momentum tensor are
\begin{align}
T_{00} &= \rho \; ,
\nonumber\\
T_{ij} &= a^2 \gamma_{ij} \cdot p \; ,
\nonumber
\end{align}
where $\rho$ and $p$ are energy density and effective
pressure, respectively.
$(00)$ and $(ij)$ components of the Einstein equations then give
\begin{subequations}
\begin{align}
H^2 &= \frac{8\pi}{3} G \rho - \frac{\kappa}{a^2}\; ,
\label{jan8-14-2}
\\
2 \dot{H} + 3 H^2 &= - 8\pi G p  - \frac{\kappa}{a^2} \; ,
\end{align}
\end{subequations}
where $H \equiv \dot{a}/{a}$ is the Hubble parameter.
A combination of these equations determines how it changes in time,
\be
\dot{H} = - 4\pi G (\rho+p) + \frac{\kappa}{a^2}
\label{jan2-14-10}
\ee
Now, one chooses
the null vector $n^\mu$ entering eq.~\eqref{jan2-14-9} as
$n^\mu = (1, a^{-1} \nu^i)$, where $\gamma_{ij}\nu^i \nu^j =1$
and finds that the NEC is equivalent in the cosmological setting to
\be 
\rho + p > 0 \; .
\nonumber
\ee
Hence, if the second, spatial curvature term in the right hand side of 
eq.~\eqref{jan2-14-10} is negaive ($\kappa < 0$, open Universe),
zero ($\kappa = 0$, spatially flat Universe) or negligible,
the Hubble parameter decreases in time. If it is negative (contraction),
it remains negative.
So, the bouncing Universe
is almost impossible. A loophole is that the bounce is possible
for closed Universe ($\kappa = +1$), provided that energy density and
pressure grow slower than $a^{-2}$ as the Universe 
shrinks\footnote{According to eq.~\eqref{jan6-14-1}, for matter
with equation of state $p=w\rho$ this requires
$w<-1/3$.}
~\cite{Starobinsky-bounce}. Note that the Penrose theorem does not
apply in the latter case, since the Cauchi hypersurface is
compact in the closed Universe (3-sphere).

Applied to the present Universe (which is spatially flat to
an excellent accuracy), the NEC implies that the Hubble
parameter cannot grow today. Observational evidence for the growing 
Hubble parameter would mean that either dark energy violates the NEC
or General Relativity is not valid at the present cosmological scales.
This would of course be highly non-trivial.

Another face of the NEC shows up through the covariant energy-momentum
conservation, $\nabla_\mu T^{\mu \nu} = 0$. In the cosmological setting
it reads
\be
\frac{d\rho}{dt} = - 3H (\rho + p) \; .
\label{jan6-14-1}
\ee
Thus, the NEC implies that energy density always decreases in
expanding Universe. Modulo the loophole mentioned above,
the Penrose theorem states that the expansion started from
a singularity -- infinite energy density, infinite expansion rate.

One more consequence of the NEC is an obstruction for the creation of
a universe in the laboratory. The question of whether one can 
{\it in principle} create a universe in the laboratory has been
raised~\cite{Berezin:1984vy,Farhi:1986ty}  
soon after the invention of inflationary theory~\cite{inflation}.
Indeed, inflation -- nearly exponential expansion of the Universe
at high expansion rate -- is capable of stretching, in a fraction of
a second, a tiny 
region of space into a region of
huge size, possibly exceeding the size of
the presently observable Universe. So, it appears at first sight
that it is
not impossible
to create artificially a region in our present Universe in which the physical 
conditions are similar to those at the onset of inflation, and then
this region would automatically expand to very large size and become 
a universe like ours. In theories obeying the NEC and within classical
General Relativity this is 
impossible~\cite{Farhi:1986ty,Berezin:1987ep} 
because of the Penrose theorem. By definition, a universe ``like ours''
is a nearly homogeneous patch in space whose size exceeds the Hubble
distance $H^{-1}$. Then the Hubble sphere is an anti-trapped surface,
and hence there had to be a singularity in the past. Since we cannot
create an appropriate singularity (and control the evolution through
any singularity), we cannot create a universe ``like ours''.
 Widely discussed ways out
are to invoke 
tunneling~\cite{Berezin:1987ea,Farhi:1989yr,Fischler:1989se,Linde:1991sk}
or other quantum 
effects~\cite{Frolov:1988vj,Guendelman:2010pr,Lukash:2013ts} and
modify gravity~\cite{Mukhanov:1991zn,Brandenberger:1993ef,Trodden:1993dm},
but it is certainly of interest to stay within General Relativity
and invoke NEC-violation instead. There were several attempts in the
latter direction~\cite{Lee:2007dh},
but many of them are problematic because of instabilities.

\begin{figure}[!tb]
\centerline{\includegraphics[width=0.5\textwidth,angle=-90]{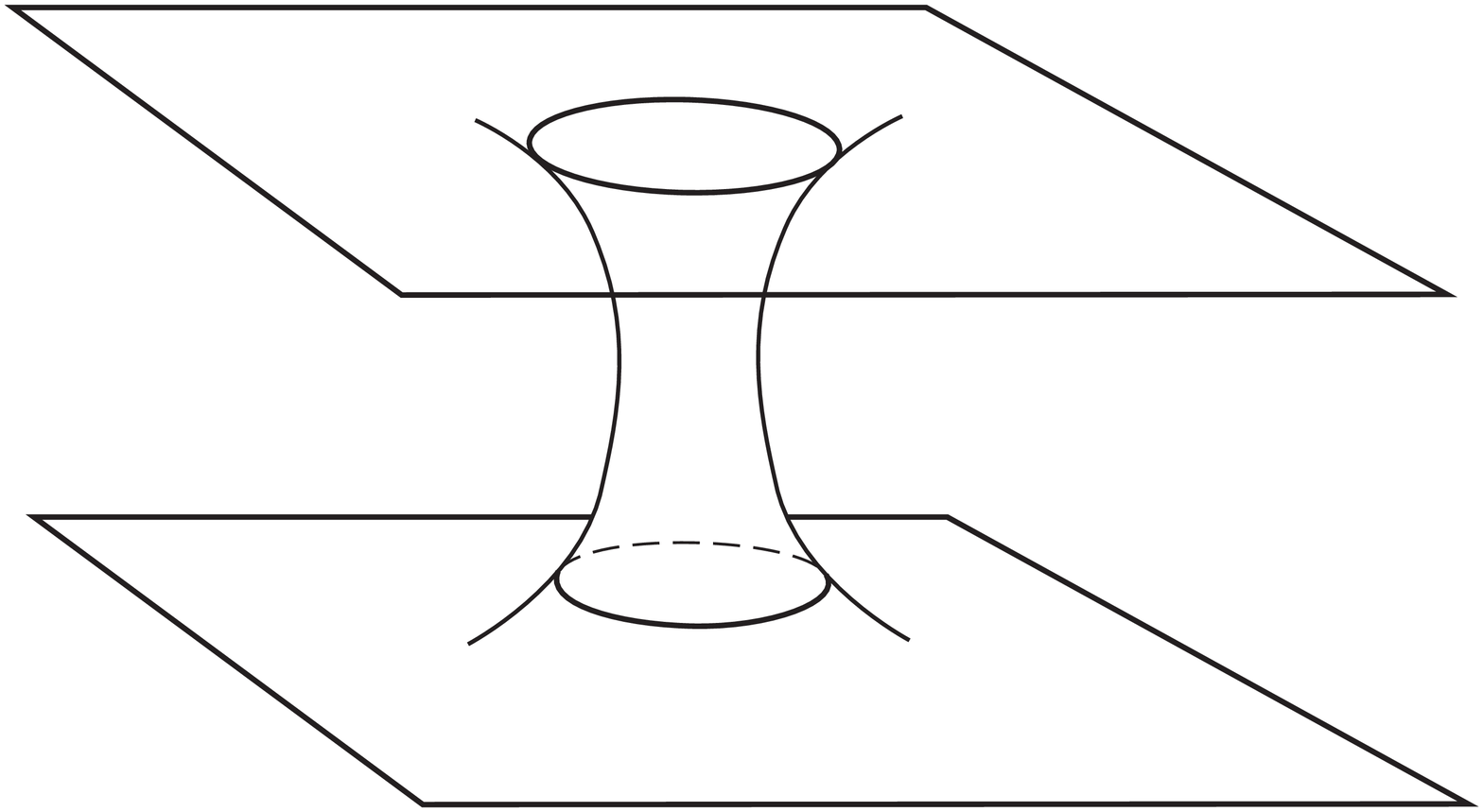}%
}
\caption{Spatial geometry of Lorentzian wormhole.
\label{fig:wormhole}
}
\end{figure}

\begin{figure}[!tb]
\centerline{\includegraphics[width=0.5\textwidth,angle=-90]{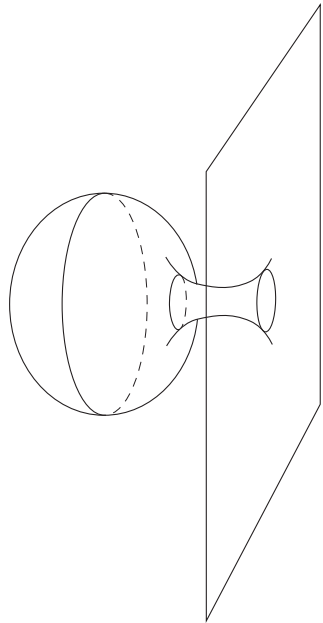}%
}
\caption{Spatial geometry of semi-closed world.
\label{fig:semiclosed}
}
\end{figure}

Finally, the NEC also forbids the existence, within General Relativity,
of throats in space, both 
static~\cite{Morris:1988cz,Morris:1988tu,Visser:1995cc} and 
time-dependent~\cite{Hochberg:1998ha}. Such a throat could join asymptotically
flat regions of space, forming a Lorenzian 
wormhole~\cite{Morris:1988cz,Morris:1988tu,Visser:1995cc,Novikov:2007zz,Shatskiy:2008us}, Fig.~\ref{fig:wormhole}. Alternatively, it could serve as a bridge
between large but finite region of space and asymptotically flat
region, forming a semi-closed world~\cite{Frolov:1988vj}, 
Fig.~\ref{fig:semiclosed}. Again, it is of interest to construct healthy
NEC-violating theories possessing wormhole solutions.

All this motivates one to search for healthy NEC-violating
theories. For a field theorist, it is natural to start with
scalar field theories. However, as we discuss in Section~\ref{robust},
solutions in theories of 
scalar fields minimally coupled to gravity and described by
Lagrangians containing first derivatives only, either obey the NEC or 
have pathologies (modulo a loophole which we briefly discuss
in Section~\ref{robust}). Because of that, one either turns to
vector fields (and, indeed, there are examples
of acceptable NEC-violating solutions in rather contrived
theories involving vector, but
not gauge, fields~\cite{Rubakov:2006pn,Libanov:2007mq}) or considers
higher-derivative  Lagrangians.
It is commonly thought, however, that theories with
Lagrangians
containing second and higher derivatives are unacceptable (unless
the higher derivative terms are treated as perturbations in the
sense of effective low energy theory),
since their field equations involve more than two derivatives,
and hence these theories have pathological degrees of freedom.
This is not the case, however: there exists a class of scalar
field theories with second-derivative Lagrangians and yet second order
field equations. These theories have been found in an unnoticed paper
by G.~W.~Horndeski~\cite{Horndeski:1974wa}, rediscovered 
in rather different context by 
 D.~B.~Fairlie, J.~Govaerts and A.~Morozov~\cite{Fairlie:1991qe}
(see Ref.~\cite{Curtright:2012gx} for a review of this approach)
and relatively recently became popular in their various
reincarnations,
such as the Dvali--Gabadadze--Porrati~\cite{Dvali:2000hr}
model in decouping 
limit~\cite{Luty:2003vm,Nicolis:2004qq}, Galileon theory~\cite{Nicolis:2008in}
and its 
generalizations~\cite{Deffayet:2009wt,deRham:2010eu,Goon-prl,Goon:2011qf,Kamada:2010qe},
k-mouflage~\cite{Babichev:2009ee}, kinetic gravity 
brading~\cite{Deffayet:2010qz,Kobayashi:2010cm,Pujolas:2011he}, 
Fab-Four~\cite{Charmousis:2011bf},
etc.
As we discuss in Section~\ref{main-1}, there exist finite number
of classes 
of second-derivative Lagrangians yielding second-derivative field
equations~\cite{Nicolis:2008in,Deffayet:2011gz,Padilla:2012dx,Sivanesan:2013tba}. At least some of these Lagrangians admit 
NEC-violation~\cite{Genesis1,Genesis2,Hinterbichler:2012fr,Hinterbichler:2012yn,Rubakov:2013kaa,Elder:2013gya}, with the NEC-violating solutions
and their neighborhoods being perfectly healthy. In 
Section~\ref{main-2}
we give a few examples
of using these theories for constructing fairly non-trivial cosmological models
and giving a proof-of-principle construction
for creating a universe in the laboratory.

We conclude in Section~\ref{conclusion} by pointing at 
some potentially problematic features of the
NEC-violating second-derivative theories
that have yet to be understood.

\section{NEC-violation and instabilities}
\label{robust}

\subsection{Tachyons, gradient instabilities, ghosts}

In this paper we are mainly interested in weak
gravity regime, 
which occurs when $M_{Pl}$ is the largest parameter in the
problem. To the lowest order this corresponds to switching off
dynamics of metric and considering other fields in Minkowski
background. 
In many cases the relevant solutions
to the fields equations are spatially homogeneous, 
and in this mini-review we stick to
his case.  We ask whether the NEC can be violated in this situation.

In a theory of one scalar field $\pi$, a spatially 
homogeneous classical solution
$\pi_c (t)$ may or may not be pathological. The pathology, if any,
shows up in the behavior of small perturbations about this
background, $\pi = \pi_c + \chi$.
Assuming that the linearized field equation for $\chi$ is second
order in derivatives, the quadratic Lagrangian for $\chi$ always
reads
\be
L_\chi^{(2)} = \frac{1}{2}U \dot{\chi}^2 - \frac{1}{2} V (\d_i \chi)^2 
-\frac{1}{2} W \chi^2 \; ,
\label{jan7-14-11}
\ee
where $U$, $V$, $W$ depend on time. Let us consider high momentum regime,
meaning that variations of $\chi$ in space and time occur at scales
much shorter than the time scale charactersitic of the background $\pi_c(t)$.
Then at a given time, the time-dependence of $U$, $V$ and $W$ can be neglected,
and there are the following possibilities:

(1) Stable background: 
\[
U>0\; , \;\;\;\;\; V>0 \; , \;\;\;\;\; W \geq 0\; .
\]
The dispersion relation
is 
\be
U \omega^2 = V {\bf p}^2 + W \; ,
\label{jan5-14-1}
\ee
which is the dispersion relation for conventional excitations,
while the energy density for perturbations 
\be
T_{00}^{(2)} =  \frac{1}{2}U \dot{\chi}^2 + \frac{1}{2} V (\d_i \chi)^2 
+ \frac{1}{2} W \chi^2
\label{jan5-14-2}
\ee
is positive, as it should. For $V<U$, the  $\chi$-waves
travel at subluminal speed, for $V=U$ they travel at the speed of light,
while for $V>U$ the $\chi$-waves are superluminal. While the superluminal
propagation is probably less of a problem, it does signal that the
theory cannot be UV-completed in a Lorentz-invariant 
way~\cite{Adams:2006sv} 
(meaning that it cannot be a low energy
theory of some Lorentz-invariant quantum theory valid at all scales), see,
however, Ref.~\cite{deRham:2013hsa} which debates this point. So, one would
like to avoid superluminality. The case $U=V$ is also potentially
problematic, since there may or may not be backgrounds {\it in a
neighborhood} of $\pi_c$, about which the perturbations are superluminal.
So, the safe case is
\be
    U > V > 0\; .
\nonumber
\ee

(1a) Special case:
\[
U > 0\; , \;\;\;\;\; V = 0 \;  .
\]
To understand how to treat this case, one thinks of the
original scalar theory as an effective field theory with UV cutoff
$\Lambda$. The Lagrangian of such a theory generically has corrections
of higher order in derivatives, wich are suppressed by powers of
$\Lambda^{-1}$ and hence are normally negligible. For $V=0$, however,
these corrections cannot be neglected, since only these corrections
give the 
terms in the Lagrangian for perturbations which involve spatial
gradients~\cite{ArkaniHamed:2003uy}. The dominant higher derivative terms
in the Lagrangian for perturbations involve second derivatives,
so the Lagrangian reads\footnote{A possible lower derivative term
$(\alpha(t)/\Lambda) \dot{\chi} (\d_i \chi)^2$, upon integration by
parts, reduces to $(\dot{\alpha}/2\Lambda)  (\d_i \chi)^2$; the pertinent
transformation reads
$(\alpha/\Lambda) \dot{\chi} (\d_i \chi)^2 \to -(\alpha/\Lambda) 
\d_i \dot{\chi} \cdot \d_i \chi =  -(\alpha/2\Lambda)\d_0(\d_i \chi)^2
\to  (\dot{\alpha}/2\Lambda)  (\d_i \chi)^2$. It is subdominant at
${\bf p}^2 \gg \dot{\alpha}{\Lambda}$, but becomes relevant at
lower momenta. It is healthy for $\dot{\alpha} < 0$.}
\be
L_\chi^{(2)} = \frac{1}{2}U \dot{\chi}^2 + \frac{1}{2\Lambda^2}
\left[ a \ddot{\chi}^2 + b \dot{\chi}^2 (\d_i \chi)^2 + c (\d_i \d_i\chi)^2
\right] \; ,
\nonumber
\ee
where we set $W=0$ for simplicity, as in the ghost condensate 
theory~\cite{ArkaniHamed:2003uy}. The dispersion relation, modulo
corrections even stronger suppressed by $\Lambda^{-1}$, is now
\be
U \omega^2 = \frac{c}{\Lambda^2} {\bf p}^4 \; ,
\nonumber
\ee
which is healthy for $c>0$. Other roots of the dispersion equation
obey $|\omega| \gg \Lambda$, so they cannot be trusted in the
low energy effective theory.

(1b) Tachyonic instability:  
\[
U>0\; , \;\;\;\;\; V>0 \; , \;\;\;\;\; W < 0\; .
\]
Formally, the
dispersion relation \eqref{jan5-14-1} yields imaginary $\omega$
for sufficiently {\it low} momenta, $V {\bf p}^2 < |W|$, so that there are
growing perturbations, $\chi \propto \exp \l \int |\omega| dt \r$
with $|\omega| \leq |W|^{1/2}$. This is indeed a problem, if 
the time scale $|W|^{-1/2}$ is much shorter than the time scale characteristic
of the background $\pi_c (t)$. In the opposite case one cannot
use the approximation of slowly varying $U(t)$, $V(t)$ and $W(t)$ 
and hence cannot
conclude that the background $\pi_c$ is unstable. Instead, 
the background is stable at short time scales, and to see what is
going on at long time scales
one
has to
perform full stability analysis. We note in passing that tachyonic
instabilities are inherent in some NEC-violating models of dark 
energy~\cite{Rubakov:2006pn,Libanov:2007mq}, 
and they may have interesting observational
consequences~\cite{Sergienko:2008tf,Libanov:2008mk}.

(2) Gradient instability:
\[
U>0\; , \;\;\;\;\; V<0 \; , \;\;\;\;\;\; \mbox{or} \;\;\;
U<0\; , \;\;\;\;\; V>0 \; .
\]
According to eq.~\eqref{jan5-14-1}, ``frequencies'' $\omega ({\bf p})$
are imaginary at high momenta, and there are perturbations that grow
arbitrarily fast. This means that the background $\pi_c$ is unstable, and 
thus not healthy. Considering the original scalar theory as an effectve
low energy theory theory valid below a certain UV scale $\Lambda$ does not
help: for consistency, the rate of variation of the background $\pi_c (t)$
must be well below $\Lambda$, while the rates of development of the 
instabilities extend up to $\Lambda$; the background is ruined at
short time scale.


(3) Ghost instability 
\[
U<0\; , \;\;\;\;\; V< 0 \;  .
\]
In {\it classical} field theory, the background is stable
against high momentum perturbations: eq.~\eqref{jan5-14-1} shows
that the frequencies are real at high momenta. Yet the background is
{\it quantum-mechanically} unstable. Indeed, the energy \eqref{jan5-14-2}
is negative at high momenta, and upon quantization the $\chi$-particles
have negative energies; they are ghosts.
The energy conservation does not forbid
pair creation from vacuum of ghosts together with
other, normal particles (say, via graviton exchange, since gravitons 
definitely
interact with $\chi$-quanta); 
vacuum is quantum-mechanically unstable.
Energies and momenta of created particles can take values up to 
the UV scale $\Lambda$  
below which one can trust the theory, so the available phase space is 
generically large, and the time scale of instability is short.
Unless $\Lambda$  
is low enough, this instability is unacceptable. So, backgrounds with
ghosts are generally considered as pathological.
We note in passing that
in Lorentz-invariant theory and for Lorentz-invariant background
$\pi_c = \mbox{const}$, the ghost instability is truly catastrophic:
if particles can be created from vacuum with some energies and momenta,
then the same, but Lorentz-boosted process is also allowed; the available
phase space is proportional to the volume of the Lorentz group,
i.e., it is infinite; the time scale of instability is infinitesimally
short. Put it differently, ghosts in the present Universe are allowed
only if Lorentz-invariance is violated in the ghost sector in such a way 
that
energies of ghost particles cannot exceed $3$~MeV~\cite{Cline:2003gs}.

The above discussion is straightforwardly generalized to a theory
with several scalar fields $\pi^I$, $I=1,\dots , N$. The Lagrangian
for perturbations $\chi^I$ is now
\be
L_\chi^{(2)} = \frac{1}{2}U_{IJ} \dot{\chi}^I \dot{\chi}^J
 - \frac{1}{2} V_{IJ} \d_i \chi^I \d_i \chi^J 
- \frac{1}{2} W_{IJ} \chi^I \chi^J \; .
\label{jan5-14-6}
\ee
and the energy density is
\be
T_{00}^{(2)} = \frac{1}{2}U_{IJ} \dot{\chi}^I \dot{\chi}^J
 + \frac{1}{2} V_{IJ} \d_i \chi^I \d_i \chi^J 
+ \frac{1}{2} W_{IJ} \chi^I \chi^J \; .
\nonumber
\ee
Barring the case of degenerate matrix $V_{IJ}$, similar to (1a) above,
the matrix $V_{IJ}$ can be diagonalized by field redefinition.
If it has negaive eigenvalue(s), the energy is unbounded from 
below~\cite{Buniy:2006xf}:
one can construct an initial configuration with $\dot{\chi}^I = 0$
with arbitrarily high momentum and $ {\bf p}^2 V_{IJ}  \chi^I \chi^J < 0$.
This is a pathological situation: there are either ghosts or
gradient instabilities, or both. For positive definite diagonal
$V_{IJ}$ one can
rescale $\chi^I$ to cast $V_{IJ}$ into unit matrix, $V_{IJ} = \delta_{IJ}$.
One can then diagonalize $U_{IJ}$ by orthogonal transformation,
so the derivative terms in the Lagrangian become
$\sum_I \left[\lambda_I (\dot{\chi}^I)^2 - (\d_i \chi^I)^2 \right]$.
If $U_{IJ}$ has negative eigenvalues $\lambda_I$, 
there are gradient instabilities. So, the requirement
of the absence of gradient instabilities
and ghosts gives the necessary condition
\be
\mbox{Stable~background:}~~~~~~~ ~~~\mbox{positive~definite}~~ U_{IJ} ,~V_{IJ} 
\; .
\label{jan5-14-7}
\ee
Whether or not there are tachyons at sufficiently low momenta
depends now on positive definiteness of $W_{IJ}$.

\subsection{Scalar theories with first-derivative Lagrangians}

As the first attempt to construct NEC-violating theory, 
one considers the Lagrangian involving
first derivatives only,
\be
L = F(X^{IJ}, \pi^I) \; ,
\label{jan5-14-5}
\ee
where 
\be
X^{IJ} = \d_\mu \pi^I \d^\mu \pi^J \; .
\nonumber
\ee
One assumes minimal coupling to gravity, then 
the energy-momentum tensor for this theory reads
\be
T_{\mu \nu} = 2 \frac{\d F}{\d X^{IJ}} \d_\mu \pi^I \d_\nu \pi^J - g_{\mu \nu} F \; .
\nonumber
\ee
Therefore, for homogeneous background
\begin{align}
T_{00} \equiv \rho &=  2 \frac{\d F}{\d X^{IJ}} X^{IJ} 
-  F
\nonumber \\
T_{11}& = T_{22} = T_{33} \equiv p = F 
\nonumber
\end{align}
and 
\be
\rho + p =  2 \frac{\d F}{\d X^{IJ}} X^{IJ} =   2 \frac{\d F}{\d X^{IJ}}
\dot{\pi}^I \dot{\pi}^J \; .
\label{jan6-14-2}
\ee
We see that NEC-violation requires that the matrix $\d F/ \d X_c^{IJ}$,
evaluated for the background $\pi_c^I$,
be non-positive definite. On the other hand, one expands the
Lagrangian \eqref{jan5-14-5} to the second order in perturbations
$\pi^I = \pi_c^I + \chi^I$ and obtains the Lagrangian for perturbations 
in the form \eqref{jan5-14-6} with
\begin{align}
U_{IJ} & = \frac{\d F}{\d X_c^{IJ}}  +
2 \frac{\d^2 F}{\d X_c^{IK} \d X_c^{JL}} \dot{\pi}_c^K \dot{\pi}_c^L
\nonumber \\
V_{IJ}  & = \frac{\d F}{\d X_c^{IJ}}  \; .
\label{jan6-14-3}
\end{align}
Thus, the stability of the background -- positive definiteness of $V_{IJ}$, 
eq.~\eqref{jan5-14-7} --
is inconsistent with NEC-violation~\cite{Buniy:2006xf}.

A loophole here is related to
the case (1a) above~\cite{Senatore:2004rj}. 
To this end, consider ghost
condensate theory with small potential 
added~\cite{Senatore:2004rj,Creminelli:2006xe},
\be
L = M^4 (X^2 - 1)^2 - V(\pi) \; ,
\nonumber
\ee
where $\pi$ is the ghost condensate field (of dimension $(\mbox{mass})^{-1}$),
$X = \d_\mu \pi \d^\mu \pi$ and $M$ is the energy scale.
In the absence of the potential, there is a solution $\pi_c = t$
for which  
$F\equiv M^4 (X^2 - 1)^2 = 0 $ and $\d F / \d X = 0$. 
This is on the borderline of
NEC-violation. Higher derivative term of appropriate sign renders this
background stable. Now, upon adding small potential $V(\pi)$
with positive slope, one makes $(\dot{\pi}_c - 1)$ slightly negative.
According to eqs.~\eqref{jan6-14-2} and \eqref{jan6-14-3}, this leads to
NEC-violation, and at the same time to the gradient instability.
However, with the higher derivative terms present,
the latter instability occurs at low momenta ${\bf p}$ only, and
can be made harmless~\cite{Creminelli:2006xe}
by careful choice of parameters and of the form of the higher derivative
corrections. This construction was used in Ref.~\cite{Creminelli:2006xe},
in particular,
to design a viable cosmological 
scenario similar to what is now called Genesis.
We discuss a less contrived Genesis model in Section~\ref{main-2}.
Also, ghost condensate idea was used to construct consistent 
bouncing Universe
models~\cite{Buchbinder:2007ad,Creminelli:2007aq}, which start from the
ekpyrotic contraction stage~\cite{Khoury:2001wf,Khoury:2001bz}.
Again, consistency of the bounce requires careful choice of parameters
in these models. We consider a simpler version of this scenario in
Section~\ref{main-2}.

\section{Second-derivative Lagrangians}
\label{main-1}

The main emphasis of this mini-review is on scalar field theories
with Lagrangians involving second derivatives, whose
equations of motion do not contain third and fourth derivatives 
nevertheless. Although nomenclature has not yet been settled, we
call them (generalized) Galileons.
We concentrate on theories of one scalar field $\pi$ in Minkowski
space and write
the Euler--Lagrange equation for a theory with the Lagrangian
$L(\pi, \d_\mu \pi, \d_\mu \d_\nu \pi)$:
\be
\frac{\d L}{\d \pi} - \partial_\mu \frac{\d L}{\d \pi_\mu}
+ \d_\mu \d_\nu \frac{\d L}{\d \pi_{\mu \nu}} = 0 \; ,
\label{jan7-14-1}
\ee
where
\be
\pi_\mu = \d_\mu \pi \; , \;\;\;\;\;\; \pi_{\mu \nu} = \d_\mu \d_\nu \pi \; .
\nonumber
\ee
Because of the last term in eq.~\eqref{jan7-14-1}, the field
equation is generically of fourth order in derivatives. However,
there are exceptions, which are precisely Galileons. 
The simplest exceptional
second derivative Lagrangian is
\be
L_{(1)} = K^{\mu \nu} (\pi, \d_\lambda \pi) \d_\mu \d_\nu \pi \; .
\label{jan7-14-2}
\ee
Off hand, the corresponding field equation is third order,
but in fact it is not. Indeed, the second term in eq.~\eqref{jan7-14-1}
gives rise to the folowing third order contribution 
\be
 - \frac{\d K^{\mu \nu} }{\d \pi_\lambda} \d_\lambda \d_\mu \d_\nu \pi \; ,
\label{jan7-14-6a}
\ee 
while the third term in eq.~\eqref{jan7-14-1} reads
\be
\d_\mu \d_\nu K^{\mu \nu} (\pi, \pi_\lambda) =  
\frac{\d K^{\mu \nu} }{\d \pi_\lambda} \d_\mu \d_\nu \d_\lambda \pi + \dots
\label{jan7-14-6b}
\ee
where omitted terms do not contain third derivatives. Hence, third order 
terms cancel out, and the field equation is second order.

It is instructive to make the following observation.
There appear to be two terms of the general form 
\eqref{jan7-14-2} with different Lorentz structure:
\be
K (\pi, X) \Box \pi \;\;\;\;\; \mbox{and} \;\;\;\;\;\;
H (\pi, X) \d^\mu \pi \d^\nu \pi \d_\mu \d_\nu \pi \; ,
\nonumber
\ee
where $\Box = \d_\lambda \d^\lambda$ and, 
as before, $X=\d_\lambda \pi \d^\lambda \pi$.
However, the second structure can be reduced to the first one by
integrating by parts (which we denote by arrow):
\be
H (\pi, X) \d^\mu \pi \d^\nu \pi \d_\mu \d_\nu \pi =
\frac{1}{2} H \d^\mu \pi \d_\mu X = \frac{1}{2} \d_\mu Q \d^\mu \pi
\to - \frac{1}{2} Q \Box \pi \; ,
\nonumber
\ee
where the function $Q(\pi, X)$ is such that $H = \d Q/\d X$. So, the only
remaining term in the Lagrangian is
\be
L_{(1)} = K_1(\pi, X) \d_\mu \d^\mu \pi \; .
\label{jan7-14-10b}
\ee
Note this term cannot be rediced by integration by parts
to any Lagrangian involving 
first derivatives only.

Let us consider a more complicated example of the Lagrangian quadratic
in the  second derivatives. There are five possible Lorentz structures:
\begin{align}
 L_{(2)} &= F_1 \d^\mu \pi \d^\nu \pi \d^\lambda \pi \d^\rho \pi
 \d_\mu \d_\nu \pi \d_\rho \d_\lambda \pi 
\nonumber\\
& + F_2 \d^\mu \pi \d^\nu  \pi  \d_\mu \d_\lambda \pi  \d_\nu \d^\lambda \pi 
\nonumber \\
& +
F_3 \d^\mu \pi \d^\nu \pi  \d_\mu \d_\nu \pi \Box \pi 
\nonumber \\ 
& +
F_4  \d_\mu \d_\nu \pi  \d^\mu \d^\nu \pi
\\
& + F_5 (\Box \pi)^2 \; ,
\nonumber
\end{align}
where $F_a = F_a (\pi, X)$, $a=1, \dots, 5$. The resulting field equation
has the following {\it fourth order} terms
\begin{align} 
& ~~~~F_1 \d^\mu \pi \d^\nu \pi \d^\lambda \pi \d^\rho \pi
 \d_\mu \d_\nu \d_\rho \d_\lambda \pi 
\nonumber\\
&+ F_2 \d^\mu \pi \d^\nu  \pi  \d_\mu   \d_\nu \Box \pi 
\nonumber \\
& +
F_3 \d^\mu \pi \d^\nu \pi  \d_\mu \d_\nu  \Box \pi 
\nonumber \\ 
& +
F_4  \Box \Box \pi
\\
& + F_5 \Box \Box \pi  \; .
\nonumber
\end{align}
We see that the fourth order terms cancel out iff
$F_1 = 0$, $F_2 = -F_3$, $F_4 = -F_5$, so that the Largagian has 
the following form
\begin{align}
L_{(2)} &=  H \d^\mu \pi \d^\nu \pi  \l
\d_\mu \d_\nu \pi \d_\lambda \d^\lambda \pi - \d_\mu \d_\lambda \pi
\d_\nu \d^\lambda \pi \r
+ K \l \d^\nu \d_\nu \pi \d_\mu \d^\mu \pi - \d^\nu \d_\mu \pi \d_\nu \d^\mu \pi 
\r 
\nonumber \\
& =
 H \d^\mu \pi \d^\nu \pi \d_\mu \d_{\left[ \nu \right.} \pi \d_{\left. \lambda \right]} 
\d^\lambda \pi + K \d^\nu \d_{\left[ \nu \right.} \pi \d_{\left. \mu\right]} \d^\mu \pi
\; ,
\label{jan7-14-5}
\end{align}
where square brackets denote anti-symmetrization (our definition
is $A_{[\mu \nu]} = A_{\mu \nu} - A_{\nu \mu}$ without numerical prefactor). 
We now understand
the reason for the cancellation of the fourth order terms
in the field equation: it happens because, e.g., $\d_\mu  
\d_{\left[ \nu \right.}  \d_{\left. \lambda \right]} \d^\mu \pi = 0$.
Again, the first term in \eqref{jan7-14-5} can be cast into the form of
the second term:
\begin{align}
 H \d^\mu \pi \d^\nu \pi  \l
\d_\mu \d_\nu \pi \d_\lambda \d^\lambda \pi - \d_\mu \d_\lambda \pi 
\d_\nu \d^\lambda \pi \r
&= \frac{1}{2} H
\l \d_\mu X \d^\mu \pi \Box \pi - 
\d_\lambda X \d_\mu \pi \d^\lambda \d^\mu \pi \r
\nonumber \\
&=  \frac{1}{2} \d_\mu Q \l  \d^\mu \pi \Box \pi - 
\d_\nu \pi \d^\nu \d^\mu \pi \r
\nonumber \\
& \to  - \frac{1}{2}  Q \l \Box \pi \Box \pi - \d_\nu \d_\mu
\pi \d^\nu \d^\mu \pi \r \; .
\nonumber
\end{align}
So, there again remains one term
\be
L_{(2)} = K_2 (\pi, X)  \l \d^\nu \d_\nu \pi \d_\mu \d^\mu \pi 
- \d^\nu \d_\mu \pi \d_\nu \d^\mu \pi \r \; .
\nonumber
\ee
Now it is straightforward to check that the third order
terms in the field equation also cancel out: the terms with
$\l \d K_2 / \d \pi_\lambda \r \d_\lambda  \l \d^\nu \d_\nu \pi \d_\mu \d^\mu \pi 
- \d^\nu \d_\mu \pi \d_\nu \d^\mu \pi \r$ cancel out automatically
in the same way as in eqs.~\eqref{jan7-14-6a}, \eqref{jan7-14-6b},
while the remaining terms like
\be
 \d^\nu K_2 (\pi, X)   \l \d_\nu  \d_\mu \d^\mu \pi 
- \d_\mu  \d_\nu \d^\mu \pi \r
\nonumber
\ee
also vanish. 

The story repeats itself in cubic and higher orders in 
the second derivatives.
The only exceptional
$n$-th order term in $D$ dimensions 
is~\cite{Nicolis:2008in,Deffayet:2011gz} 
\be
L_{(n)} = K_n (\pi, X) \d^{\mu_1} \d_{\left[ \mu_1 \right.} \pi
\dots \d^{\mu_n} \d_{\left. \mu_n \right]} \pi \; .
\label{jan8-14-10}
\ee
The fact that the corresponding field equation is second order is
checked trivially; the proof that no other terms exist is not so
simple~\cite{Deffayet:2011gz} . Note that $L_{(n)}$ can be written as
\be
L_{(n)} = \frac{1}{(D-n)!} 
K_n(\pi, X) \epsilon^{\nu_1 \dots \nu_{D-n} \mu_1 \dots \mu_n}
 \epsilon_{\nu_1 \dots \nu_{D-n} \lambda_1 \dots \lambda_n} \d_{\mu_1}
\d^{\lambda_1} \pi \dots \d_{\mu_n}
\d^{\lambda_n} \pi \; .
\label{jan8-14-14}
\ee
Indeed, any antisymmetric tensor $A_{\mu_1  \dots \mu_n}$
can be written as
\be
A_{\mu_1  \dots \mu_n} = \epsilon_{\nu_1 \dots \nu_{D-n} \mu_1 , \dots, \mu_n}
B^{\nu_1 \dots \nu_{D-n}} \; ,
\label{jan8-14-11}
\ee
where
\be
B^{\nu_1 \dots \nu_{D-n}} = \frac{1}{n! (D-n)!} 
\epsilon^{\nu_1 \dots \nu_{D-n} \mu_1 , \dots, \mu_n} A_{\mu_1  \dots \mu_n}
\label{jan8-14-12}
\ee
is the dual tensor. The expression in the right hand side of
eq.~\eqref{jan8-14-10} is antisymmetric in both  upper and lower
indices. Applying the transformation~\eqref{jan8-14-11}, \eqref{jan8-14-12}
to upper indices and to lower indices separately, one arrives at 
the form \eqref{jan8-14-14}.
 
Note that there are $(D+1)$ allowed classes of Lagrangians, if one counts 
the class without second derivatives
\be
L_{(0)} = K_0 (\pi, X) \; .
\label{jan7-14-10a}
\ee
In particular, there are five classes in four dimensions.
A general Galileon Lagrangian is a sum of all these terms.

This completes the discussion of the
exceptional theories of one scalar field, Galileon,
in Minkowski space. Theories with {\it multiple} scalar fields are considered
in Refs.~\cite{Padilla:2012dx,Sivanesan:2013tba}
(see also Refs.~\cite{Padilla:2010de,Padilla:2010ir,Hinterbichler:2010xn}). 
The minimal generalization
of $L_{(1)}$ to  {\it curved} space-time is simple:
\be
L_{(1)} = K_1 (\pi, X) \nabla_\mu \nabla^\mu \pi \; ,
\nonumber
\ee
where $X= g^{\mu \nu} \d_\mu \pi \d_\nu \pi$; it is straightforward to
check that the resulting field equation is still second order.
The energy-momentum tensor, and hence the Einstein equations,
are second oreder in derivatives as well.
The generalizations of $L_{(2)}$ and higher order Lagrangians
are, on the other hand,
non-trivial~\cite{Horndeski:1974wa,Deffayet:2009wt,Deffayet:2011gz}.
 Finally, we note that
some Galileon Lagrangians 
have an interesting interpretation as describing a three-brane evolving
in five-dimensional space-time~\cite{deRham:2010eu,Goon-prl}.

\section{Examples of NEC-violation}
\label{main-2}

In this Section we consider an example of simple NEC-violating solution
and its use for constructing rather non-trivial cosmological
scenarios. We also discuss the possibility of creatng a universe in the
laboratory by employing the 
Galileon models of Section~\ref{main-1}. Our set of
illustrations is of course personal and by no means complete.

\subsection{Rolling background}
\label{main2-1}

The analysis is particularly simple in models exhibiting scale invariance,
\be
\pi (x) \to \pi^\prime(x) = \pi(\lambda x) + \ln \lambda \; .
\label{may4-13-5}
\ee 
It is sufficient for our purposes to consider the
Lagrangian involving only 
the terms $L_{(0)}$ and $L_{(1)}$, eqs.~\eqref{jan7-14-10a}
and \eqref{jan7-14-10b}. Theories of this general form
have been studied in detail in Refs.~\cite{Deffayet:2010qz,Kobayashi:2010cm,Pujolas:2011he} and named kinetic gravity braiding.
We write in the 
scale-invariant case and in Minkowski space
\be
L_\pi = F (Y) \e^{4\pi} + K(Y) \Box \pi \cdot \e^{2\pi} \; ,
\label{may3-13-1}
\ee
where
\be
Y = \e^{-2\pi} (\d \pi)^2 \; , \;\;\;\;\;\;\;\; (\d \pi)^2 \equiv
\d_\mu \pi \d^\mu \pi \; ,
\label{mar5-13-2}
\ee
and $F$ and $K$ are yet unspecified functions.
Assuming that $K$ is analytic near the origin, we  set 
\be
K(Y=0)= 0
\label{may6-13-1} \; .
\ee 
Indeed, upon integrating by parts, a constant part of $K$
can be 
absorbed into the $F$-term in
eq.~\eqref{may3-13-1}. 
We will need the expression for the energy-momentum tensor.
To this end, we consider minimal coupling to the metric, i.e., set
$Y = \e^{-2\pi} g^{\mu \nu} \d_\mu \pi \d_\nu \pi$ and $\Box \pi =
\nabla^\mu \nabla_\mu \pi$ in curved space-time. To calculate the 
energy-momentum 
tensor, we note that in curved space-time,
the $K$-term in $\sqrt{-g} L_\pi$ 
can be written, upon integrating by parts, as
$\sqrt{-g} g^{\mu \nu} \d_\mu \pi \d_\nu \l K \e^{2\pi} \r$.
Then the
variation with respect to $g^{\mu \nu}$ is straightforward, and we get
\begin{align*}
T_{\mu \nu} &= 2 F^\prime \e^{2\pi} \d_\mu \pi \d_\nu \pi - g_{\mu \nu} F \e^{4\pi}
\nonumber
\\& + 2 \Box \pi \cdot K^\prime  \d_\mu \pi \d_\nu \pi -
\d_\mu \pi \cdot \d_\nu \l K \e^{2\pi} \r-
\d_\nu \pi \cdot \d_\mu \l K \e^{2\pi} \r
+ g_{\mu \nu} g^{\lambda \rho} \d_\lambda \pi \d_\rho \l K \e^{2\pi} \r \; .
\end{align*}
This expression is valid in curved space-time as well.

In what follows we consider homogeneous backgrounds in Minkowski
space, $\pi = \pi (t)$.
For a homogeneous field,
the field equation reads
\begin{align}
& 4 \e^{4\pi} F + F^\prime \e^{2\pi}\left(- 6 \dot{\pi}^2
- 2\ddot{\pi} \right) - 
 2 \e^{2\pi} \dot{\pi}  F^{\prime \prime} \dot{Y}
\nonumber\\
 & + K \e^{2\pi} \left( 4 \dot{ \pi}^2 + 4 \ddot{\pi} \right)
+ 4 \e^{2\pi} \dot{ \pi} K^\prime \dot{Y}
 + K^{\prime \prime} \dot{Y} \left( -2  \dot{ \pi}^3 \right)
 + K^\prime \left( - 12\dot{ \pi}^2  \ddot{\pi} + 4 \dot{ \pi}^4
\right)
=0 \; ,
\label{mar5-13-3}
\end{align}
while the energy density and pressure are
\begin{subequations}
\begin{align}
\rho &= \e^{4\pi} Z
\label{may3-13-11}
\\
p &= \e^{4\pi} \l F - 2YK - \e^{-2\pi} K^\prime \dot{\pi} \dot{Y} \r \; ,
\end{align}
\end{subequations}
where
\[
Z= -F + 2Y F^\prime - 2YK + 2Y^2K^\prime \; .
\]
It is straightforward to see that for 
$\dot{\pi} \neq 0$, eq.~\eqref{mar5-13-3}
is equivalent to energy conservation,
$\dot{\rho} = 0$.

It is instructive to calculate the quadratic Lagrangian for 
perturbations about the homogeneous background. It has the form
\eqref{jan7-14-11} with
\begin{subequations}
\begin{align}
U &= \mbox{e}^{2\pi_c} \l F^\prime + 2 Y F^{\prime \prime}
- 2K + 2 Y K^\prime + 2 Y^2 K^{\prime \prime} \r
= \e^{2\pi_c} Z^\prime \; ,
\label{may4-13-2}
\\
V &= \e^{2\pi_c}\l F^\prime - 2K + 2YK^\prime -
2 Y^2 K^{\prime \prime} \r + \l 2K^\prime + 2Y K^{\prime \prime}
\r \ddot{\pi}_c \; .
\label{may3-13-2}
\end{align}
\end{subequations}
We will not need the general expression for $W$.
Note that $U$ is proportional to the derivative with respect to $Y$
of the same function $Z$ that determines the energy density,
eq.~\eqref{may3-13-11}. 

With $F(0) =0$, the theory admits a constant solution
$\pi_c = \mbox{const}$, $Y=0$ and $T_{\mu \nu} = 0$. In the absence
of other forms of energy, this solution corresponds to Minkowski
space. Equations~\eqref{may4-13-2}, \eqref{may3-13-2}
show that the Minkowski background is stable for
$F^\prime (0) > 0$, and that perturbations travel with the speed of light
(recall that we set $K(0)=0$).
This is easy to understand: it follows from eq.~\eqref{may3-13-1}
that perturbations about constant $\pi_c$ are governed by the first term there,
and $L^{(2)} = \e^{2\pi_c} F^\prime (0) (\d \chi)^2$, which is the Lagrangian for
a massless scalar field. In the neighbourhood of the Minkowski background,
i.e., for small $\d \pi_c$, perturbations are not 
superluminal~\cite{Rubakov:2013kaa} 
provided that $K^\prime (0) = 0$, $F^{\prime \prime} (0) > 0$.

In a wide range of the functions $F$ and $K$,
eq.~\eqref{mar5-13-3} admits also a rolling solution,
\be
\e^\pi = \frac{1}{\sqrt{Y_*}(t_* - t)} \; ,
\label{may4-13-1}
\ee
where $t_*$ is an arbitrary constant.
For this solution
$Y =Y_* = \mbox{const}$,
and $Y_{*}$ is determined from equation 
\be
Z(Y_*) \equiv -F + 2Y_{*} F^\prime - 2 Y_{*}  K + 2 Y^2_{*} K^\prime  = 0 \; ,
\label{feb27-13-1}
\ee
where $F$, $F^\prime$, etc., are evaluated at $Y=Y_*$.
For this solution one has $T_{00}= \rho =0$ and
\be
p = \frac{1}{Y_*^2(t_* - t)^4} \l F - 2Y_* K \r \; .
\label{mar3-13-2}
\ee
Thus, the rolling background violates the NEC, provided that
\be
\mbox{NEC-violation:}~~~~~~~~~~~~~~~~
2 Y_{*}K - F >0 \; .
\label{mar3-13-6}
\ee

The quadratic Lagrangian
for perturbations \eqref{jan7-14-11} reduces in this background to
\be
L^{(2)} = \frac{A}{Y_{*}(t_* - t)^2} [\dot{\chi}^2 -(\d_i \chi)^2] + 
 \frac{B}{Y_{*}(t_* - t)^2} \dot{\chi}^2 +  
\frac{C}{Y_{*}^2(t_* - t)^4} \chi^2 \; ,
\label{mar3-12-1}
\ee
where 
\begin{align*}
A&=\e^{-2\pi_c} V = F^\prime - 2 K + 4 Y_{*} K^\prime 
\\
B &= \e^{-2\pi_c}(U-V)=
 2 Y_{*}F^{\prime \prime} - 2 Y_{*} K^\prime + 2 Y^2_{*} K^{\prime \prime}
\\
C &= 8 F - 12 Y_{*} F^\prime + 8 Y_{*}^2 F^{\prime \prime}
+ 8 Y_{*} K - 8 Y_{*}^2 K^\prime + 8 Y^3_{*} K^{\prime \prime}
\end{align*}
are time-independent coefficients. 
As a cross check, one can derive from the latter Lagrangian
the equation for homogeneous perturbation
$\chi(t)$ about the rolling background and see that $\chi = \d_{t} \pi_c =
(t_* - t)^{-1}$ obeys this equation, as it should. Indeed, making use of
eq.~\eqref{feb27-13-1} one finds that the
coefficients of $\dot{\chi}^2$
and $\chi^2$ in eq.~\eqref{mar3-12-1} are related in a simple way,
\[
4(A+B) = C/Y_{*} \; .
\]
Hence, homogeneous perturbation obeys a universal equation
\[
- \frac{d}{dt} \l \frac{\dot{\chi}}{(t_* - t)^2} \r
+ 
 4\frac{\chi}{(t_* - t)^4} = 0 \; ,
\]
whose solutions are $\chi = (t_* - t)^{-1}$ and $\chi= (t_* - t)^4$.
This shows that the rolling background is an attractor
and that it is stable against low momentum 
perturbations:
the growing perturbation $\chi = (t_* - t)^{-1}
\cdot \chi_0({\bf x})$ with slowly varying  $\chi_0({\bf x})$
can be absorbed into slightly inhomogeneous time shift.

Let us consider the stability of the rolling background and
subluminality of the perturbations about it. 
The spatial gradient term  in \eqref{mar3-12-1}
has correct (negative) sign provided that
\be
\mbox{No~gradient~instability:}~~~~~~~~~~~~~~~~
 A=F^\prime - 2 K + 4 Y_{*} K^\prime > 0 \; .
\label{mar3-13-4}
\ee
The speed of perturbations about the rolling background is smaller
than the speed of light, if the coefficient of $\dot{\chi}^2$
is greater than that of $-(\d_i \chi)^2$, i.e.,
\be
\mbox{Subluminality:}~~~~~~~~~~~~~~~~
B=2Y_* F^{\prime \prime} - 2Y_* K^\prime + 2Y^2_{*} K^{\prime \prime} > 0 \; .
\label{mar3-13-5}
\ee
We require that this inequality holds in strong sense,
then the perturbations about the rolling solution are strictly
subluminal, and hence the perturbations about backgrounds
neighbouring the rolling solution
are subluminal as well. 
When both inequalities \eqref{mar3-13-4} and  \eqref{mar3-13-5}
are satisfied,
there are no ghosts either.
The conditions \eqref{mar3-13-6}, \eqref{mar3-13-4} and 
\eqref{mar3-13-5} together with eq.~\eqref{feb27-13-1} can be satisfied
at $Y=Y_*$
by a judicious choice of the functions $F$ and $K$ in the neighbourhood
of this point, so that the NEC-violation is stable and subluminal.
This can be seen as follows. Equation~\eqref{feb27-13-1}
can be used to express $F(Y_*)$ in terms of $F^\prime (Y_*)$, $K(Y_*)$ and
$K^\prime (Y_*)$, namely, $F=2Y_* F^\prime - 2 Y_* K + 2 Y_*^2 K^\prime$.
Then the inequalities  \eqref{mar3-13-6}, \eqref{mar3-13-4} are satisfied,
provided that $2 K - 4Y_* K^\prime < F^\prime < 2 K - Y_* K^\prime$, 
which is possible
for positive $K^\prime$. The condition \eqref{mar3-13-5} can be satisfied
by an appropriate choice of $F^{\prime \prime}$ and $K^{\prime \prime}$.

To make contact with existsing studies, we note that
a particular Lagrangian of the type \eqref{may3-13-1} considered in 
Ref.~\cite{Genesis2} is
\be
L_\pi = -f^2 \e^{2\pi} (\d \pi)^2 + \frac{f^2}{2\Lambda^3} (1+\alpha) (\d \pi)^4
+ \frac{f^3}{\Lambda^3} (\d \pi)^2 \Box \pi \; ,
\label{jan8-14-1}
\ee
which corresponds to
\be
F = - f^2 Y + \frac{f^2}{2\Lambda^3} (1+\alpha) Y^2 \; , \;\;\;\;\;\;
K= \frac{f^3}{\Lambda^3} Y \; .
\nonumber
\ee
Here the parameters $f$ and $\Lambda$ have dimension of mass, the parameter
$\alpha$ is dimensionless. The solution to eq.~\eqref{feb27-13-1} is
\be
Y_* = \frac{2}{3(1+\alpha)} \frac{\Lambda^3}{f} \; .
\nonumber
\ee
One requires that the energy scale $\sqrt{Y_*}$ associated with this
solution is lower than $\Lambda$, which is interpreted as the UV cutoff
scale. This gives
\be
f \gg \Lambda \; .
\label{jan8-14-3}
\ee
From eq.~\eqref{mar3-13-6} one finds that
the background $Y=Y_*$ violates the NEC iff
\be
2 Y_{*}K - F = 2 f^2 Y_* \frac{3+\alpha}{1+\alpha} > 0 \; ,
\nonumber
\ee
while the stability and subluminality conditions, eqs.~\eqref{mar3-13-4}
and \eqref{mar3-13-5}, give
\be
A = \frac{3-\alpha}{3(1+\alpha)}f^2 > 0 \; ,
\;\;\;\;\;\;\;\; B = \frac{4\alpha}{3(1+\alpha)} f^2 > 0 \; .
\nonumber
\ee
All these conditions are satisfied for~\cite{Genesis2} 
\be
0 < \alpha < 3 \; .
\nonumber
\ee
Note that the case $\alpha = 0$ corresponds to luminal propagation of
perturbations about the background $Y=Y_*$. In fact, in this case the 
theory \eqref{jan8-14-1} is invariant 
under conformal symmetry~\cite{Nicolis:2008in,Genesis1}. However,
the case $\alpha = 0$ is problematic, since there are backgrounds
in the neighborhood of $Y=Y_*$ about which the propagation of
perturbations is superluminal~\cite{Genesis2}. Note also that
the Lagrangian \eqref{jan8-14-1} does not admit stable Minkowski
background, since $F^\prime (0) < 0$. Conformally invariant theory
with stable Minkowski background and subluminal propagation
about the solution $Y=Y_*$ and in its neighborhood was constructed in
Ref.~\cite{Hinterbichler:2012yn} building upon Ref.~\cite{deRham:2010eu},
and goes under the name DBI conformal Galileon theory. 

To end up this Section, let us consider~\cite{Rubakov:2013kaa} 
the structure of the
configuration space $(\pi, \dot{\pi})$ of spatially homogeneous
Galileons  
in arbitrary Galileon theory with scale invariance
\eqref{may4-13-5}. The Lagrangian may contain all terms discussed in
Section~\ref{main-1}.
We pointed out above  that for 
$\dot{\pi} \neq 0$, the field equation is 
equivalent to energy conservation,
$\dot{\rho} = 0$.
This is not an accident. The Noether theorem states that
the Noether energy-momentum tensor
(which coincides with the metric energy-momentum tensor
for the scalar field minimally coupled to gravity) obeys
\[
\d_\mu T^{\mu}_{\nu} = - (\mbox{E.O.M.})\cdot \d_\nu \pi \; ,
\]
where  $(\mbox{E.O.M.})$ stands for the equation of motion.
Therefore, 
the equation of motion
for spatially homogeneous $\pi = \pi (t)$
is 
\be
(\mbox{E.O.M.}) = - \frac{1}{\dot{\pi}} \dot{\rho} \; .
\label{may5-13-1}
\ee
Since the field equation is second order, $\rho = \rho(\pi, \dot{\pi})$ 
does not contain $\ddot{\pi}$ and higher derivatives, and by
scale invariance it has the form 
\be
\rho = \e^{4\pi} \, Z(Y) \; , 
\nonumber
\ee
where
$Y= \dot{\pi}^2 \, \exp(-2\pi) $, cf. eq.~\eqref{mar5-13-2},
and $Z$ is a model-dependent
function. 
Now we can understand in more general terms that the rolling background 
with $Z=0$ and $\dot{\pi} > 0$
is an attractor in the class of homogeneous solutions. To this end, we
use the conservation of energy $\dot \rho = 0$
and write for any homogeneous solution
\be
\e^{4\pi} Z =  \mbox{const}  \; .
\label{apr19-13-2}
\ee
As $\pi$ increases, $|Z|$ decreases, so the solution tends to a
configuration with $Z \to 0$. The configuration space of homogeneous 
Galileons with $\dot{\pi} > 0$
is thus divided into basins of attraction of solutions with $Z=0$.

We also pointed out above that the coefficient $U$ entering the
quadratic action for perturbations is proportional to $Z^\prime$.
This is not an accident 
either. 
To see this,
let us again use eq.~\eqref{may5-13-1}
valid for any homogeneous Galileon.
It follows from this equation that
the equation of motion for homogeneous perturbation about
the background $\pi_c (t)$ reads
\[
 - \frac{1}{\dot{\pi}_c} \frac{\d \rho}{\d \dot{\pi}_c} \ddot{\chi} + \dots 
= 0 \; ,
\]
where omitted terms do not contain $\ddot{\chi}$. Hence, 
the Lagrangian
for  perturbations has the form
\[
L^{(2)} =
\frac{1}{2\dot{\pi}_c}  \frac{\d \rho}{\d \dot{\pi}_c} \dot{\chi}^2
= \dots = \e^{2\pi_c} Z^\prime (Y)  \dot{\chi}^2 = \dots\; ,
\]
where omitted terms do not contain $\dot{\pi}$. We conclude that
$\rho = \e^{4\pi_c} \, Z(Y_c)$, $U= \e^{2\pi_c} Z^\prime(Y_c)$
for any point in the configuration
space of homogeneous Galileon
$(\pi_c, \dot{\pi}_c)$ in any scale-invariant Galileon theory.

Recall finally that a point  in the configuration
space $(\pi_c, \dot{\pi}_c)$, 
at which $U<0$, is unstable: there is either ghost or gradient
instability among perturbations about this point.
The above results therefore mean that any path in the space
of homogeneous configurations $(\pi , \dot{\pi})$ that connects
two zero energy attractor solutions, $Z=0$, passes through
an unstable region: indeed, $Z^\prime$ is negative somewhere
at this path. This property creates
difficulties in using scale-invariant
Galileons, as we discuss later on. Here we note that 
it implies that 
there is no evolution without pathologies
that connects the Minkowski and rolling backgrounds, even if this
evolution is driven by a source (provided that this source does not couple
to $\dot{\pi}$). 

The above analysis heavily uses scale invariance. Once one gives up
scale invariance, this analysis and its conclusions are no longer
valid. In particular, evolution from nearly Minkowski regime to
rolling regime can occur without pathologies~\cite{Elder:2013gya}.

\subsection{Genesis scenario}
\label{sec:Genesis}

As the first example of utilizing the solution discussed in 
Section~\ref{main2-1}, let us consider Galilean
Genesis~\cite{Genesis1} -- a cosmological scenario alternative
to inflation(see also
Refs.~\cite{Genesis2,Liu:2011ns,Liu:2012ww,Hinterbichler:2012fr,Hinterbichler:2012yn,Liu:2013xt}). 
One assumes that at early times $t\to -\infty$, the 
space-time is
Minkowskian, energy and pressure vanish, the Universe is empty.
At that time the only relevant form of matter is the Galileon field $\pi$
described by the Lagrangian~\eqref{may3-13-1} (other Galileon Lagrangians
are considered in Refs.~\cite{Liu:2011ns,Hinterbichler:2012yn,Nishi:2014bsa}
with fairly similar results). Once the conditions
\eqref{mar3-13-6}, \eqref{mar3-13-4} and \eqref{mar3-13-5} are satisfied,
the solution $Y=Y_*$ is stable and violates the NEC. At the initial stage
of evolution, i.e., at large enough $(t_* - t)$, energy density and pressure
are small, and
one can make use of the
perturbation theory in $G\equiv M_{Pl}^{-2}$.
Equation~\eqref{jan2-14-10}
with $\kappa =0$ determines the Hubble parameter, and to the lowest non-trivial
order in $M_{Pl}^{-2}$ one makes use of the Minkowski expressions for
energy density, $\rho=0$, and
pressure, eq.~\eqref{mar3-13-2},
\be
p =  - \frac{P}{(t_* - t)^4} \; ,
\label{jan10-14-2}
\ee
where $P = (2Y_* K -F)Y_*^{-2}$.
One finds
\be
H = \frac{4\pi P}{3M_{Pl}^2 (t_* - t)^3} \; .
\nonumber
\ee
Equation~\eqref{jan8-14-2} is then used to find the energy density
to the first order in $M_{Pl}^{-2}$:
\be
\rho = \frac{3}{8\pi} M_{Pl}^2 H^2 = \frac{3\pi}{8} 
\frac{P^2}{M_{Pl}^2 (t_*- t)^6} \; .
\nonumber
\ee
We see that as the field
$\pi_c$ evolves, energy density builds up,
and the cosmological expansion gets accelerated. The weak gravity
approximation (expansion in $M_{Pl}^{-2}$) is valid when $\rho \ll p$,
i.e.,
\be
(t_*-t)^2 \gg \frac{P}{M_{Pl}^2} \; .
\label{jan8-14-5}
\ee
The parameter $P$ may be large: in the example with the 
Lagrangian~\eqref{jan8-14-1} one has $P \sim f^3/\Lambda^3 \gg 1$
in view of eq.~\eqref{jan8-14-3}. Still,
if $P$ is not exceedingly large, the weak gravity regime holds
almost to the Planck scale. 

As a cross check, let us consider the field equation for 
homogeneous $\pi$ in
expanding spatially flat Universe. It reads
\begin{align}
& 4 \e^{4\pi} F + F^\prime \e^{2\pi}\left(- 6 \dot{\pi}^2
- 2\ddot{\pi} \right) + 
   4 F^{\prime \prime} \l - \dot{\pi}^2 \ddot{\pi} +  \dot{\pi}^4\r
\nonumber\\
 & + 4 K \e^{2\pi} \left(  \dot{\pi}^2 +  \ddot{\pi} \right)
-4 K^\prime \l \dot{\pi}^2 \ddot{\pi} + \dot{\pi}^4 \r
 + 4 \e^{-2\pi}
K^{\prime \prime} \l - \dot{\pi}^4 \ddot{\pi} + \dot{\pi}^6 \r
\nonumber \\
& - 6 H \e^{2\pi} \dot{\pi} F^\prime 
 + 12 H \e^{2\pi} \dot{\pi} K
\nonumber \\
&- 6 K^{\prime} \l 2 H \dot{\pi}^3 + 2 H \dot{\pi} \ddot{\pi}
+ \dot{H} \dot{\pi}^2 + 3 H^2 \dot{\pi}^2 \r
 + 12 \e^{-2\pi} H K^{\prime \prime} \l -\dot{\pi}^3 \ddot{\pi} + \dot{\pi}^5 \r
=0 \; .
\end{align}
We see that gravitational corrections here are small provided that
$H \ll \dot \pi$, which again gives the condition \eqref{jan8-14-5}.
Discussing weak gravity regime is sufficient for our purposes, but
of course one can follow the evolution after the end of this
regime, with gravity effects fully accounted for. This is done 
in Ref.~\cite{Genesis1} in the model \eqref{jan8-14-1} with
$\alpha = 0$. 

So far we have seen that the theory admits a cosmological scenario
in which the Universe starts empty and Minkowskian and evolves into
the stage of rapid expansion and high energy density. This evolution is
precisely the Genesis epoch. There are two other ingredients in
the Genesis scenario. First, at some late stage the Galileon energy density
should be converted into heat, and the standard hot epoch should begin.
A possible mechanism of ``defrosting'' is suggested in 
Ref.~\cite{LevasseurPerreault:2011mw}. At the end of ``defrosting''
stage, whatever it is, the Galileon should settle to its Minkowski
value, $\dot{\pi} = 0$. In the scale-invariant Galileon theory this
is problematic because of our observations in the end of 
Section~\ref{main2-1}. The violation of scale invariance at
``defrosting'' can probably
cure this problem.

The second ingredient is a mechanism of the generation of 
density perturbations,
responsible in the end for CMB anisotropies and structure formation.
These perturbations are Gaussian (or nearly Gaussian) random field
with nearly flat power spectrum.
Perturbations in the Galileon field itself cannot do the job~\cite{Genesis1}.
A simple extension of the Galileon theory can, however, work quite 
well~\cite{Genesis1} (see Refs.~\cite{Liu:2011ns,Liu:2012ww} 
for alternative proposals).
One insists on scale invariance at the Genesis epoch and adds a new field
$\theta$ which trivially transforms under scale transformations,
$\theta (x) \to \theta (\lambda x)$. By scale invariance, the kinetic term
in its Lagrangian is 
\be
L_\theta = \frac{1}{2} \e^{2\pi} (\d \theta)^2 \; .
\nonumber
\ee
If other interactions of the new field are negligible at the Genesis epoch,
the Lagrangian in the rolling background \eqref{may4-13-1} is
\be
L_\theta = \frac{1}{2} \frac{1}{Y_* (t_* - t)^2} (\d \theta)^2 \; .
\nonumber
\ee
This coincides with the Lagrangian of a scalar field minimally coupled
to gravity, evolving at inflationary epoch with the Hubble parameter
$\sqrt{Y_*}$, if one identifies $t$ with conformal time at inflation.
Thus, one borrows the well-known result of the inflationary theory:
vacuum fluctuations of the field $\theta$ develop into Gaussian
random field with the power spectrum
\be
{\cal P}_{\delta \theta} = \frac{Y_*}{(2\pi)^2} \; .
\label{jan9-14-1}
\ee
The field perturbations $\delta \theta$, which are
entropy fluctuations at the Genesis epoch,  
are assumed to be reprocessed into adiabatic perturbations
sometime after the Genesis epoch by, say, curvaton~\cite{Linde:1996gt}
or modulated decay~\cite{Dvali:2003em} mechanism.
The adiabatic perturbations $\zeta$
inherit the properties of perturbations
$\delta \theta$ (modulo non-Gaussianities that may be produced 
in the process of
conversion of entropy to adiabatic perturbtion); 
in particular, their power spectrum
is ${\cal P}_\zeta = \mbox{const} \cdot {\cal P}_\theta$. The spectrum
\eqref{jan9-14-1} is flat; small tilt, 
required by observations~\cite{Hinshaw:2012aka,Ade:2013zuv},
can emerge due to weak explicit breaking of scale invariance 
(cf. Ref.~\cite{Osipov:2010ee}).

To conclude this Section, we note that the Genesis scenario,
especially its version with conformal Galileon, is an example of what
is now called (pseudo-)conformal 
cosmology~\cite{Rubakov:2009np,Genesis1,Hinterbichler:2011qk,Hinterbichler:2012fr}.
In general terms, this class of scenarios assumes that the Universe
is initially effectively Minkowskian, and matter is in conformally
invariant state. Then conformal invariance is spontaneously
broken by rolling background similar to
\eqref{may4-13-1}. The mechanism of the generation of density perturbations
is similar to one just discussed. Conformal scenario makes a number
of model-independent predictions which potentially distinguish it
from inflation. These include non-Gaussianities and statistical
anisotropy of scalar perturbations~\cite{Libanov:2010nk,sabir}. Another property
is the absence of tensor perturbations.

\subsection{Bouncing Universe}

Galileon theories can also be used to construct models of 
bouncing 
Universe~\cite{Genesis1,Qiu:2011cy,Easson:2011zy,Cai:2012va,Cai:2013vm,Osipov:2013ssa,Qiu:2013eoa,Koehn:2013upa}.
Before we discuss a concrete model of this sort, let us make the following
comment. Contracting Universe can easily become strongly inhomogeneous and
anisotropic
because of the Belinsky--Lifshits--Khalatnikov 
phenomenon~\cite{Lifshitz:1963ps}. This makes a problem for consistency
of the entire bouncing scenario.
A way to cure this problem is to assume that the dominant matter at the
contracting stage has super-stiff equation of state, 
$p>\rho$~\cite{Erickson:2003zm}. This is what one generically
calls ekpyrotic Universe~\cite{Khoury:2001wf,Khoury:2001bz}. We 
discuss 
this point in Appendix B. Note that for matter with the equation of
state $p=w\rho$,  $w=\mbox{const}$, 
eq.~\eqref{jan6-14-1} gives $\rho \propto a^{-3 (1+w)}$, 
and then one finds from eq.~\eqref{jan8-14-2} with $\kappa=0$ that the
scale factor evolves as 
\be
a(t) \propto |t|^\alpha \; , \;\;\;\;\;\; t<0 \; ,
\nonumber
\ee
where
\be
\alpha = \frac{2}{3(1+w)} \; .
\nonumber
\ee
Super-stiff equation of state, $w>1$, thus corresponds to
\be
\alpha < \frac{1}{3} \; .
\label{noin-11*}
\ee

An example of super-stiff matter is a scalar field with the negative
exponential potential,
\be
L_\phi = \frac{1}{2} \d_\mu \phi \d^\mu \phi - V(\phi) \; , \;\;\;\;\;\;\;
V(\phi) = - V_0 \,\e^{\phi/M} \; ,
\label{noin-add7}
\ee
where $V_0$ and $M$ are positive parameters. The equation for the
homogeneous field 
$\phi(t)$ and the Friedmann equation\eqref{jan8-14-2}
 have the following solution,
\be
a(t) = |t|^\alpha \; , \;\;\;\;\;\;\;
\phi(t) = \mbox{const} - 2M \log |t| \;, \;\;\;\;\;\;\;
V[\phi(t)] = -\frac{2M^2 (1-3\alpha)}{t^2} \; , \;\;\;\;\; t < 0 \; ,
\label{noin-12+}
\ee
where
\be
\alpha =16 \pi\, \frac{M^2}{M_{Pl}^2} \; .
\label{noin-12*}
\ee
This is an attractor in the case of collapse.
According to
\eqref{noin-11*} and \eqref{noin-12*}, the effective
equation of state is
indeed super-stiff,  $w \gg 1$, for
$M \ll M_{Pl}$. Note that the energy density is positive and increases
as the Universe collapses,
\[
\rho = \half \dot{\phi}^2 + V(\phi) = \frac{6M^2 \alpha}{t^2} \; .
\]
This leaves open the possibility that the 
potential $V(\phi)$ becomes positive
at large $\phi$, 
and the field moves out of
the negative potential at some late epoch.

It is worth noting that for $M\ll M_{Pl}$
this solution is always in the weak gravity 
regime similar to that
studied in Section~\ref{sec:Genesis}. In the weak
gravity  limit one neglects gravity
in the field equation for $\phi$ and obtains the solution in
Minkowski space
\be
\phi (t) =  M \log \l \frac{2M^2}{t^2 V_0} \r \; , \;\;\;\;\;\;
V(t) = - \frac{2M^2}{t^2} \; .
\label{jan10-14-1}
\ee
The energy density  vanishes in this limit, while pressure is
\be
p = \frac{1}{2} \dot{\phi}^2 - V = \frac{4 M^2}{t^2} \; .
\label{jan10-14-4}
\ee
The weak gravity approximation is valid at all times for
$M \ll M_{Pl}$.


While one can construct models with just one scalar field, in which
ekpyrotic contraction ends up in a 
bounce~\cite{Cai:2012va,Cai:2013vm,Qiu:2013eoa,Koehn:2013upa},
it is a lot simpler~\cite{Osipov:2013ssa} to 
extend the model \eqref{noin-add7} by adding
a Galileon field with the Lagrangian \eqref{may3-13-1},
so that the total matter Lagrangian is
\be
L = L_\pi + L_\phi \; .
\ee
In the weak gravity limit, the fields $\phi$ and $\pi$ do not
interact with each other, the Galileon rolls as in eq.~\eqref{may4-13-1},
while $\phi (t)$ is given by \eqref{jan10-14-1}. Energy density is zero, and
pressure is the
sum of \eqref{jan10-14-2} and \eqref{jan10-14-4}:
\be
p = \frac{4 M^2}{t^2} - \frac{P}{(t_* - t)^4} \; .
\ee
The Hubble parameter is found from
eq.~\eqref{jan2-14-10}:
\be
 H = - \frac{16 \pi M^2}{M_{Pl}^2 |t|} + \frac{4\pi P}{3M_{Pl}^2 (t_* - t)^3}
\; .
\ee
At early times, the field $\phi$ dominates and the Universe contracts
($H<0$),
later on the Galileon takes over, at least for $t_* < 0$, the 
contraction terminates ($H=0$, bounce), 
the expansion epoch begins and proceeds
like in the Genesis scenario ($H>0$). It is straightforward to see that
the bounce indeed occurs in the weak gravity regime,
$H \ll \dot{\pi}$, provided that the following mild inequality holds,
$|t_*| \gg P^{1/2} M^2/M_{Pl}^3$, $t_* < 0$ (the case $t_* > 0$ is considered
in Ref.~\cite{Osipov:2013ssa} with the result that 
the bounce is always there, but it
occurs not necessarily in the weak gravity regime). 

To make this toy model more realistic, one modifies the potential $V(\phi)$
at large $\phi$ and adds the potential to the Galileon to ensure
that the cosmological constant vanishes at late times. Depending on parameters,
the system may or may not enter the late time inflationary 
regime~\cite{Osipov:2013ssa}. The ingredients discussed in 
the end of Section~\ref{sec:Genesis} have to be present in this model as well.

\subsection{Creating a universe in the Laboratory}
\label{sec:creation}

Our last example is an attempt to design a model for the creation of
a universe in the laboratory~\cite{Rubakov:2013kaa}. The idea is to
construct initial condition in a Galileon-type theory such that inside
some large sphere the field $\pi$ is nearly homogeneous and behaves like
at the initial stage of
Genesis, whereas outside this sphere this field tends to a
constant and space-time is asymptotically Minkowskian.
For this initial data, 
the energy density and pressure are initially small everywhere and
the entire space-time is nearly Minkowskian, so that the required
field configuration can in principle be prepared in the laboratory. 
As the field $\pi (t, {\bf x})$ 
evolves from this initial state according to
its equation of motion, the energy density inside the large sphere
increases, space undergoes accelerated expansion there, and the region 
inside the sphere eventually 
becomes a man-made universe. Outside this sphere the energy density
remains small and asymptotes to zero at large distances;
the space-time is always asymptotically Minkowskian.

It is tempting to implement this idea 
in a simple way, by
considering the initial field
$\pi (t, {\bf x})$ which slowly varies
in space and interpolates between the rolling solution 
\eqref{may4-13-1}
inside the large
sphere and Minkowski vacuum $\d \pi = 0$ at spatial infinity.
By slow variation in space we mean that the spatial derivatives of
$\pi$ are negligible compared to temporal ones, so that at each 
point in space $\pi$ evolves in the same way as in the homogeneous
case.

An advantage of this quasi-homogeneous approach is its simplicity; 
a disadvantage is that it actually does not work in the class of
scale-invariant models of Section~\ref{main2-1}. The obstruction
comes from the property discussed in the end of Section~\ref{main2-1}:
if the evolution of $\pi$
is effectively homogeneous everywhere, then the analysis of 
Section~\ref{main2-1} applies, and since
 $Z(Y)$ vanishes
both inside the large sphere (Genesis region)
and far away from it (Minkowski region), there is a region in between
where $Z^\prime < 0$ and the system is unstable.

One way to get around this obstruction would be to insist on
slow spatial variation of the initial field configuration but
give up the prescription that the field inside the large sphere
is in the Genesis regime \eqref{may4-13-1}. Instead, one would consider
the field with non-zero energy density inside the sphere,
so that there exists a smooth and stable configuration
that interpolates, as $r$ increases, between this field and the
asymptotic Minkowski vacuum. This can hardly lead to the creation
of a universe, however, since, as we discussed in Section~\ref{main2-1},
the Minkowski
point $Y = 0$ is an attractor, and the field
in the interior of the sphere will relax to it.

Other possibilities are to consider field configurations with
non-negligible spatial gradients or give up scale-invariance of
the action (the latter possibility has been 
successfully explored in Ref.~\cite{Elder:2013gya}
in the cosmological context). 
In either case
the above no-go 
agrument would be
irrelevant, but the analysis would be more complicated.
It is simpler to follow
another route, and complicate the model instead.

To this end, one allows 
the functions $F$ and $K$ to depend explicitly on spatial
coordinates. This can be the case 
if there is another field, call it $\varphi$, which 
determines the couplings entering these functions, and this field
acts as time-independent background, $\varphi = \varphi({\bf x})$.
In this case one can consider a field configuration
$\pi (t, {\bf x})$ which at any point in space is approximately
given by the rolling solution \eqref{may4-13-1}, but with $Y_*$
depending on ${\bf x}$.  
One prepares the background $\varphi ({\bf x})$ in such a way
that $Y_*({\bf x})$ is constant inside the large sphere
(to evolve into a man-made universe) and gradually aproaches zero
as $r \to \infty$. It is straightforward to check that with an appropriate
choice of the functions $F(Y;\varphi)$, $K(Y;\varphi)$, this 
construction does not encounter pathologies anywhere.

\begin{figure}[!htb]
\centerline{\includegraphics[width=0.5\textwidth,angle=-90]{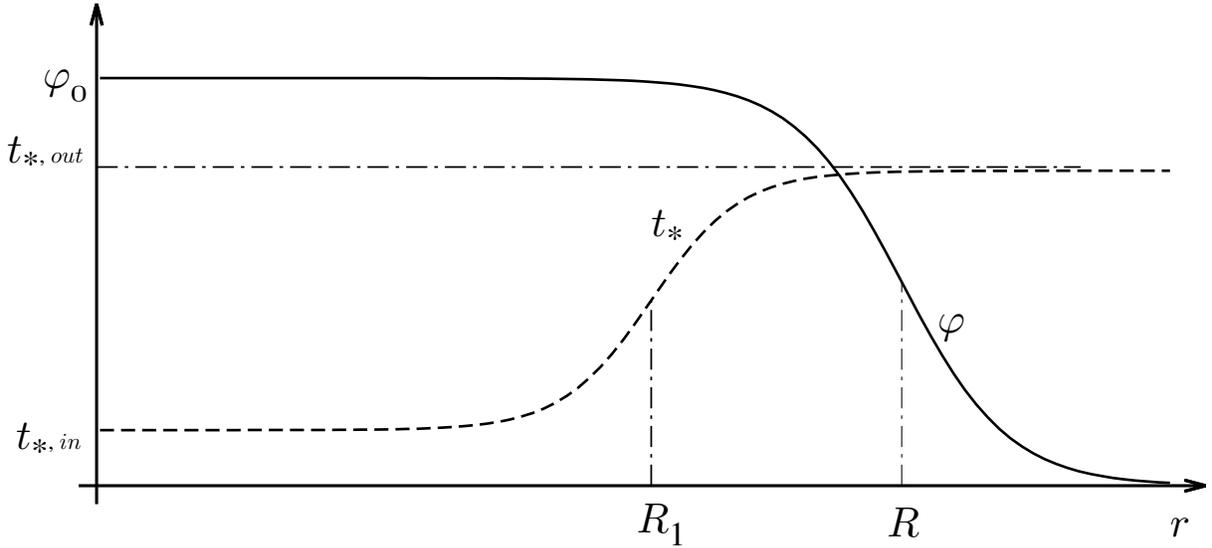}%
}
\caption{The set up. Dashed and solid lines show $t_* (r)$ and $\varphi (r)$,
respectively. The behaviour of the
function $Y_{*} (r) = Y_{*} (\varphi (r))$ is similar 
to that of $\varphi (r)$.
\label{fig1}
}
\end{figure}

Let us now sketch a concrete construction. 
Let us assume that
the field $\varphi$ is a usual scalar field which has two
vacua, $\varphi = 0$ and $\varphi = \varphi_0$. We prepare
a spherical configuraion of this field with  $\varphi = \varphi_0$
inside a sphere of large enough radius $R$ and  $\varphi = 0$
outside this sphere, see Fig.~\ref{fig1}. 
We assume for definiteness 
that there is a source for the field
$\varphi$ that keeps this configuration static.
Let $L \ll R$ be the thickness of the wall separating
the two vacua; $L$ is also kept time-independent by the source.
We require that the mass of this ball is small enough, so that
$R \gg R_s$, where $R_s$ is the Schwarzschild radius.
The mass is of order $\mu^4 R^2 L$, where $\mu$ is the mass scale 
characteristic of the field $\varphi$. Hence, the latter requirement
reads $\mu^4 R L \ll M_{Pl}^2$. For small enough $\mu$ both $R$ and $L$
can be large.

Let the function $Y_{*} (\varphi)$ 
be such that $Y_{*} (0) = 0$ and  $Y_{*} (\varphi_0) = Y_0$.
We prepare the initial configuration of $\pi$ at $t=0$ in such a way
that it initially evolves as
\be
\e^{\pi} = \frac{1}{\sqrt{Y_0}t_* (r) - 
\sqrt{Y_{*}(r)}t} \; ,
\label{may9-13-1}
\ee
where we allow the parameter $t_*$ in \eqref{may4-13-1} to vary in space,
and choose a convenient parametrization. We choose $t_* (r)=t_{*, in}$ 
inside a somewhat smaller sphere of radius $R_1 < R$
(but $R_1 \sim R$) and 
$t_*(r)=t_{*, out} \gg t_{*, in}$ at $r > R_1$ (hereafter
subscripts {\it in} and {\it out}
refer to the regions $r < R_1$ and $r> R_1$, respectively),
as shown in Fig.~\ref{fig1},  with the transition region
of, say, the same thickness $L$.
We take $t_{*, out} \ll L$, then the
characteristic time scales are smaller than the smallest length scale $L$
inherent in the set up,
so the spatial devivatives of $\pi$ are indeed
negligible compared to 
the time derivatives. This ensures that
the field $\pi$ is in the quasi-homogeneous regime.
As $r \to \infty$, we have $Y_{*}(r) \to 0$ and $t_* \to \mbox{const}$, so
 the field $\pi$ tends to the
Minkowski vacuum $\pi = \mbox{const}$.

At the initial stage of evolution, 
pressure inside the sphere of radius $R_1$ is
\[
p_{in} = \frac{M^4}{Y_0^2 (t_{*, in} - t)^4} \; ,
\]
where $M$ is the mass scale characteristic of the field $\pi$.
We require that 
$p_{in} R^3/M_{Pl}^2 \ll R $,
then the gravitational potentials are small everywhere,
and gravity is initially
in the linear regime. Thus, we impose a constraint
\be
    \frac{M^4 R^2}{Y_0^2 t_{* , in}^4} \ll M_{Pl}^2 \; ,
\label{may7-13-1}
\ee
which is consistent with the above conditions for
$M \ll M_{Pl}$ and $Y_0 \gtrsim M^2$. In complete analogy with
Section~\ref{sec:Genesis},
the Hubble parameter
inside the sphere of radius $R_1$ shortly after the beginning of
evolution is 
\be
H_{in} = \frac{4 \pi M^4}{3 M_{Pl}^2 Y_0^2 (t_{* , in} - t)^3} \; .
\label{may9-13-2}
\ee
In view of \eqref{may7-13-1} and $t_{* ,in} \ll R$, the Hubble
length scale is  large for some time, $H^{-1} \gg R$.
This is true also at $r > R_1$, so
there are no anti-trapped surfaces initially. 

As $t$ approaches $t_{*, in}$, pressure in the Genesis region $r< R_1$ 
increases,
and the Hubble length shrinks there to $R_1 \sim R$. The 
anti-trapped surfaces get formed inside the sphere of radius $R_1$,
a new universe gets created and enters the Genesis regime there. 
This occurs when $H_{in} \sim R^{-1}$, i.e., at time $t_1$ such that
\[
(t_{* , in} - t_1) \sim \l \frac{M^4 R}{M_{Pl}^2 Y_0^2 } \r^{1/3}\; .
\]
Note that at that time the energy density 
$\rho_{in} \sim M_{Pl}^2 H_{in}^2$ is still relatively small,
\[
\frac{\rho_{in}}{p_{in}} \sim \l \frac{M^4}{Y_0^2 R^2 M_{Pl}^2} \r^{1/3} 
\ll 1 \; .
\]
This implies that at time $t_1$, space-time is locally nearly
Minkowskian. Another manifestation of this fact is that the scale
factor is close to 1:
\be
a_{in}(t_1) = 1 + \frac{2\pi M^4}{3 M_{Pl}^2 Y_0^2 (t_{* , in} - t_1)^2}
\label{may16-13-1}
\ee
where the correction to 1 is of order $\rho_{in}/p_{in}$. Hence,
our approximate solution \eqref{may9-13-1}, \eqref{may9-13-2} is legitimate.

Since $t_{* , out} \gg t_{* , in}$, the field $e^{\pi}$ at time $t_1$
is still small
at $r>R_1$, and the Hubble length scale exceeds $R$ there.
Gravity is still weak at $r>R_1$, so it is consistent to assume that
the configuration of $\varphi$ is not modified by that time.
Note also that
a black hole is not formed by then either.

This completes the construction of the initial configuration 
and the analysis of the early
epoch of a man-made universe. We make contact between this analysis
and the general results of Ref.~\cite{Berezin:1987ep}
in Appendix C.

Of course, the construction discussed here is merely a sketch.
To make the scenario complete, one has to
specify the way to design the configuration of the field $\varphi$
and keep it static (or consider an evolving field $\varphi$ instead).
Also, one has to understand
the role of spatial gradients.
Finally, one would like to trace the dynamics of the system to longer
times, with gravity effects included,
and see what geometry develops towards the end of the Genesis
epoch occuring at $r<R_1$. In particular, it is of interest to see whether
a black hole gets formed.

\section{Conclusion}
\label{conclusion}

The theories of (generalized) Galileons offer an interesting possibility
of consistent and controllable NEC-violation. Still, there remain
open issues. One of them is the danger of superluminality.
While the background we consider in Section~\ref{main2-1} may be
safe in this respect, it is not impossible that other backgrounds
are sick, especially when gravity generated by some other matter is
relevant. An example of this sort is given in Ref.~\cite{Easson:2013bda}.
The superluminality issue is tightly related to the possibility
of UV completion~\cite{Adams:2006sv}. Another issue is the stability
against the radiative corrections. While the simplest Galileon theories
possess enough symmetries to guarantee the stability, generic Galileon
Lagrangians \eqref{jan8-14-10} do not. There are also largely unexplored
areas where NEC-violating theories may make surprizes, 
like black hole thermodynamics~\cite{Dubovsky:2006vk},
absence/existence of closed time-like curves~\cite{Hawking:1991nk}
and naked singularities~\cite{Curtright:2012uz}, etc.

Of course, the most intriguing question is whether NEC-violating
fields exist in Nature. Needless to say, no such fields have been
discovered. The situation is not entirely hopeless, however:
we may learn at some point in future
that the Universe went through the bounce
or Genesis epoch, and that will be an indication that NEC-violation
indeed took place in the past.

\vspace{0.5cm}

The author is indebted to S.~Demidov, D.~Levkov, M.~Libanov, 
I.~Tkachev and M.~Voloshin
for helpful discussions
and S.~Deser,   Y.-S.~Piao and  A.~Vikman for useful
correspondence. This work has been supported in part
by the grant of the President of the Russian Federation
NS-5590.2012.2 and the Ministry of Education and Science contract
8412.

\section{Appendix A}

Let us consider general spherically symmetric metric
which we choose in diagonal form
\be
ds^2 = N^2 dt^2 - a^2 dr^2 - R^2 (d\theta^2 + \sin^2 \theta d\varphi^2) \; ,
\label{jan2-14-1a}
\ee
where $N=N(t,r)$, $a=a(t,r)$, $R=R(t,r)$.
Our purpose is to show that trapped sphere is such that
$R(t,r(t))$ decreases along outgoing null geodesic, for which 
$r$ increases. 

 The formal definition of
a trapped sphere is that 
\be
  \nabla_\mu l^\mu < 0
\nonumber
\ee
for a vector $l^\mu = d x^\mu /d\lambda$ tangent to outgoing
radial null geodesic, where $\lambda$ is affine parameter.
Vector $l^\mu$ is null, 
\be
g_{\mu \nu} l^\mu l^\nu =0 \; ,
\label{jan2-14-2}
\ee
and obeys the geodesic equation 
\be
\frac{d l^\mu}{d\lambda} + \Gamma^\mu_{\nu \rho} l^\nu l^\rho = 0 \; .
\label{jan2-14-3}
\ee
For metric \eqref{jan2-14-1a},
eq.~\eqref{jan2-14-2} gives
\be
l^0 = u(t) \; , \;\;\;\;\;\;\; l^r = u(t) \frac{N(t,r(t))}{a(t, r(t))} \; ,
\nonumber
\ee
where we have chosen to parametrise the geodesic by time $t$,
so that the null world line is $(t, r(t), 0, 0)$; 
the sign of $l^r$ corresponds to outgoing geodesic. The normalization 
factor $u(t)$ is to be 
determined from eq.~\eqref{jan2-14-3}. To find this factor,
we write $d l^\mu/d \lambda = d l^\mu/dt \cdot l^0$ and obtain the
for the 
0-th component of eq.~\eqref{jan2-14-3}:
\be
 \frac{du}{dt} + \Gamma^0_{00} u + 2 \Gamma^{0}_{0r} u \frac{N}{a}
+ \Gamma^0_{rr} u \l \frac{N}{a} \r^2 = 0 \; .
\label{jan2-14-5}
\ee
The relevant
Christoffel symbols are
\be
 \Gamma^{0}_{00} = \frac{\dot{N}}{N} \; ,
\;\;\;\;\; \Gamma^0_{0r} = \frac{N^\prime}{N}  \; ,
\;\;\;\;\; \Gamma^0_{rr} = \frac{a \dot{a}}{N^2} \; ,
\nonumber
\ee
where dot and prime denote {\it partial} derivatives,
and Christoffel symbols entering eq.~\eqref{jan2-14-5}
are to be taken at $r=r(t)$. Thus, the function $u(t)$
obeys
\be
\dot{u} +\l \frac{\dot{N}}{N} + 2 \frac{N^\prime}{a} + \frac{\dot{a}}{a} \r u
=0 \; ,
\label{jan2-13-7}
\ee
where, again, the terms in parenthesis are partial derivatives
evaluated at $r=r(t)$. As a cross check, one writes the $r$-component 
of the geodesic equation \eqref{jan2-14-3},
\be
\frac{d}{dt} \left[u \frac{N(t, r(t))}{a(t, r(t))} \right]
+ \Gamma^r_{00} u + 2 \Gamma^{r}_{0r} u \frac{N}{a}
+ \Gamma^r_{rr} u \l \frac{N}{a} \r^2 = 0 \; .
\label{jan2-13-6}
\ee
One makes use of
\be
 \Gamma^{r}_{00} = \frac{N N^\prime}{a^2} \; ,
\;\;\;\;\; \Gamma^0_{0r} = \frac{\dot{a}}{a}  \; ,
\;\;\;\;\; \Gamma^0_{rr} = \frac{a^\prime}{a} \; 
\nonumber
\ee
and
\be
 \frac{dr(t)}{dt} = \frac{N}{a} \; ,
\nonumber
\ee
and finds that eq.~\eqref{jan2-13-6} coincides with
eq.~\eqref{jan2-13-7}.

We now calculate
\begin{align}
\nabla_\mu l^\mu &= \frac{1}{\sqrt{-g}}\d_\mu \l \sqrt{-g} l^\mu \r
=\frac{1}{aNR^2}\left[
\d_0 (aNR^2 u) + \d_r \l aNR^2 u \frac{N}{a} \r \right]
\nonumber\\ 
&= \dot{u} + \l \frac{\dot{a}}{a} + \frac{\dot{N}}{N} + 2 \frac{\dot{R}}{R}
+ 2 \frac{N^\prime}{a} + 2 \frac{R^\prime}{R} \frac{N}{a} \r u 
\nonumber
\end{align}
Using eq.~\eqref{jan2-13-7} to eliminate $\dot{u}$, we arrive at
\be
\nabla_\mu l^\mu = 2 \l \frac{\dot{R}}{R}u + \frac{R^\prime}{R} \frac{N}{a} u
\r = 2 l^\mu \d_\mu R \; .
\label{jan2-14-8}
\ee
Thus, the trapped surface
 is indeed such that
$R(t,r(t))$ decreases along outgoing null geodesic.

As an example, for contracting spatially flat Universe we have
$a=a(t)$, $R=a(t)r$, and the right hand side of eq.~\eqref{jan2-14-8}
is negative for $r > -1/\dot{a}$;
a sphere of radius
$R =ar > |H|^{-1}$ is trapped surface. 
By time reversal, a sphere of radius
$R > |H|^{-1}$ in expanding Universe is anti-trapped surface.

\section{Appendix B}

Let us briefly discuss why the contracting Universe
gets strongly inhomogeneous and anisotropic if the dominant matter
obeys $p<\rho$, and why, on the contrary, it becomes more 
isotropic in the course of contraction in the opposite case.
Let us consider a simplified version of anisotropic Universe
which is
descibed by the homogeneous anisotropic metric
\be
ds^2 = dt^2 - a^2 (t) \cdot \sum_{a=1}^{3} \mbox{e}^{2 \beta_a\l
  t\r}\, 
e_i^{(a)} e_j^{(a)} dx^i dx^j \; ,
\label{noin-9**}
\nonumber
\ee
where $e_i^{(a)}$ are three linear independent vectors which are constant in 
time. We assume for simplicity that these vectors
are orthogonal to each other (the dynamics is a lot more complicated
in the general situation, but this turns out to be largely
irrelevant from our viewpoint, see a comment below). 
The function
$a(t)$ is chosen in such a way that
\be
\sum_a \beta_a = 0 \; ;
\label{noin-add1}
\ee
in other words, $\mbox{det} g_{ij} = a^6$. 
The Einstein equations give
\begin{subequations}
\begin{align}
\l \frac{\dot{a}}{a} \r^2 &= \frac{1}{6} \sum_a \dot{\beta}^2_a
+ \frac{8\pi}{3} G \rho\;, 
\label{noin-9+}
\\
\ddot{\beta}_a + 3 \frac{\dot{a}}{a} \dot{\beta}_a &= 0 
\; .
\label{noin-9*}
\end{align}
\end{subequations}
Equation \eqref{noin-9*} gives
\be
\dot{\beta}_a = \frac{d_a}{a^3} \; ,
\label{noin-10*}
\ee
and in view of \eqref{noin-add1}, the constants $d_a$ obey
$\sum_a d_a = 0$. Then Eq.~\eqref{noin-9+} becomes
\be
\l \frac{\dot{a}}{a} \r^2 = \frac{1}{6a^6} \sum_a d^2_a
+ \frac{8\pi}{3} G \rho \; . 
\label{noin-10**}
\ee
This equation shows that the overall contraction rate (the rate
at which $\mbox{det} g_{ij}$ decreases) is determined at small $a$
by the anisotropy rather than matter, provided that
$\rho$ increases slower than
$a^{-6}$. For metric 
\eqref{noin-9**} covariant energy conservation still gives
eq.~\eqref{jan6-14-1} with $H=\dot{a}/a$,
so the latter property holds for
$p<\rho$. Therefore, one can set
$\rho=0$ late at the collapsing stage, and the system of equations
\eqref{noin-10*}, \eqref{noin-10**} has the Kasner
solution
\begin{align}
a(t) &= |t|^{1/3} \; , \;\;\;\;\;
\beta_a = d_a \log |t|
\nonumber
\\
\sum_a d_a &= 0  \; , \;\;\;\;\;\;\;\;\;
\sum_a d_a^2 = \frac{2}{3} \; .
\nonumber
\end{align}
Hence, the anisotropy increases as the Universe collapses.
In the general case when the vectors
$e^{(a)}_i$ are not orthogonal to each other, this regime continues
for finite time, and then the values of 
the parameters
$d_a$ change in a rather abrupt manner~\cite{LandauLifshitz2}. 
The vectors
$e^{(a)}_i$ change too. This change occurs infinitely many times in
the limit 
$t\to 0$. This corresponds to the chaotic anisotropic collapse.

These results show that 
the Universe is very anisotropic before the
bounce. In fact, the processes we described occur independently in 
Hubble-size regions and are very different in each of them because of their 
chaotic properties, so the Universe becomes strongly inhomogeneous too. 
This picture remains valid after the bounce,
at least in the framework of the classical theory.
Strong inhomogeneity of the Universe after the bounce
is inconsistent with the smallness of the primordial cosmological
perturbations, so the entire bounce scenario is up in the air. 

To solve this problem, one invokes matter with
super-stiff equation of state
$p=w\rho$, $w>1$. Its energy density behaves as
$\rho \propto a^{-3(1+w)}$, so it increases faster than  $a^{-6}$. 
The second term in the right hand side of
Eq.~\eqref{noin-10**} dominates, the scale factor decreases as
$a(t) \propto |t|^\alpha$ with $\alpha < 1/3$, see eq.~\eqref{noin-11*}.
It then follows from
Eq.~\eqref{noin-10*} that the parameters
$\beta_a$ tend to constants as
$ t \to 0$. If the Universe is nearly homogeneous at the early stages 
of collapse, and anisotropy is not strong, then the Universe becomes
more and more homogeneous in the process of contraction, see
details in
Ref.~\cite{Erickson:2003zm}.

\section{Appendix C}

We show in this appendix that the results of
Section~\ref{sec:creation} are in agreement with the general
results of Ref.~\cite{Berezin:1987ep}. 

{\it Definition~\cite{Berezin:1987ep}. Let the metric have the
form~\eqref{jan2-14-1a}.
R-region is a region where normal vectors $R_\mu = \d_\mu R$ 
to hypersurfaces
$R=\mbox{const}$ are spacelike, $g^{\mu \nu} R_\mu R_\nu < 0$.
Since $g^{\mu \nu} R_\mu R_\nu = N^{-2} \dot{R}^2 - a^{-2} R^{\prime 2} < 0$,
there is no place in an R-region where $R^\prime = 0$, so the
sign of $R^\prime$ is one and the same in the entire R-region.
An  R-region where $R^\prime > 0$ is called $R_+$-region,
while an  R-region where $R^\prime < 0$ is called $R_-$-region.
T-region is a region where normal vectors $R_\mu$ to hypersurfaces
$R=\mbox{const}$ are timelike, $g^{\mu \nu} R_\mu R_\nu > 0$.
There $\dot{R}$ is non-zero everywhere. Hence  the sign of
$\dot{R}$ is the same everywhere. A  T-region where $\dot{R} > 0$ 
is called $T_+$-region,
while a  T-region where $\dot{R} < 0$ is called $T_-$-region.
$T_+$- and $T_{-}$-regions are regions of expansion and contraction,
respectively.}

Let us consider the model of Section~\ref{sec:creation}.
In the above nomenclature, the whole space is initially $R_+$-region.
At time $t_1$, a $T_+$-region appears. One of its boundaries moves
towards smaller $r$, and another moves towards larger $r$. 
One of the results of Ref.~\cite{Berezin:1987ep} is that for
$\rho + p < 0$ ($\beta < 0$ in nomenclature of  Ref.~\cite{Berezin:1987ep}),
the boundary  between an inner $R_+$ region and $T_+$ region
is necessarily space-like. Let us check that our geometry is
consistent with this result.

In our case, $N=1$, $a \approx 1$ (see \eqref{may16-13-1})
and $R = a (r,t) r$. The boundary between the left $R_+$-region
and $T_+$-region is determined by $\dot{a} r = a$, i.e.,
\[
r - H^{-1} = 0 \; .
\]
The normal to this hypersurface is the vector 
\[
\l \frac{\dot H}{H^2} \, , 1\, , 0 \, ,  0 \r \; ,
\]
which is timelike, since
\[
\frac{\dot H}{H^2} \sim \l H (t_{* , in} - t) \r^{-1}
\sim \frac{M_{Pl}^2 Y_0^2 (t_{* , in} - t)^2}{M^4}  \gg 1 \; .
\]
Hence, the hypersurface separating the $R_+$- and $T_+$-regions is
spacelike, in agreement with the general result of
Ref.~\cite{Berezin:1987ep}.

The outer boundary of the $T_+$-region may be in principle either
spacelike or
timelike~\cite{Berezin:1987ep}. For the same reason as above, it is
actually spacelike in our case.

\end{document}